\documentclass[10pt,aps,pra,twocolumn,superscriptaddress]{revtex4-2}

\usepackage[utf8]{inputenc}
\usepackage[english]{babel}

\usepackage{amsmath,amssymb,amsfonts,amsbsy,amsthm}
\usepackage{mathtools}
\usepackage{physics}
\usepackage{braket}
\usepackage{bm}
\usepackage{bbm}
\usepackage{dsfont} 

\usepackage{graphicx}
\usepackage{xcolor}
\usepackage[colorlinks]{hyperref}

\begin{document}

\title{Exact analysis of AC sensors based on Floquet time crystals}

\author{Andrei Tsypilnikov}
\affiliation{Instituto de F\'isica, Universidade Federal Fluminense, Av. Gal. Milton Tavares de Souza s/n, Gragoat\'a, 24210-346 Niter\'oi, Rio de Janeiro, Brazil}
\author{Matheus Fibger}
\affiliation{Instituto de F\'isica, Universidade Federal Fluminense, Av. Gal. Milton Tavares de Souza s/n, Gragoat\'a, 24210-346 Niter\'oi, Rio de Janeiro, Brazil}
\author{Fernando Iemini}
\affiliation{Instituto de F\'isica, Universidade Federal Fluminense, Av. Gal. Milton Tavares de Souza s/n, Gragoat\'a, 24210-346 Niter\'oi, Rio de Janeiro, Brazil}

\begin{abstract}
	We discuss the behavior of general Floquet time crystals (FTCs), including prethermal ones, in closed systems acting as  AC sensors. We provide an analytical treatment of their quantum Fisher information (QFI) dynamics, which characterizes the ultimate sensor accuracy. By tuning the direction and frequency of the AC field, we show how to induce transitions resonantly between macroscopic paired cat states in the FTC sensor. This allows for robust Heisenberg scaling precision (QFI \(\sim N^2 t^2\)) for exponentially long times in the system size. The QFI dynamics exhibit, moreover, a characteristic step-like structure in time due to the eventual dephasing along the cat subspaces. The behavior is discussed for various initial sensor preparations, including ground states and low- and high-correlated states. Furthermore, we examine the performance of the sensor along the FTC phase transition, with the QFI capturing its critical exponents. Our findings are presented for both linear and nonlinear response regimes and illustrated for a specific FTC based on the long-range interacting LMG model.
\end{abstract}

\maketitle

\section{Introduction}
\label{sec:introduction}

The non-equilibrium dynamics of many-body quantum systems offer a fascinating way to explore novel phases of matter, revealing emergent phenomena that cannot be explained by conventional equilibrium descriptions. In strongly interacting systems, their collective dynamics can give rise to new behaviors including many-body localization, prethermalization, and dynamical phase transitions \cite{Heyl_2018,abaninColloquiumManybodyLocalization2019}. These exotic phases not only deepen our understanding of quantum matter but also hold promise for next-generation quantum technologies. In particular, their quantum nature and collective behavior can enhance sensing precision beyond classical and non-interacting limits, opening new avenues for high-precision metrology \cite{montenegroReview2025a,Degen_review,pezzeQuantumMetrologyNonclassical2018}.

A notable class of these emerging phases is that of time crystals, which are non-equilibrium, many-body phases of matter that spontaneously break time-translation symmetry and exhibit long-range temporal order. Initially proposed by Wilczek in isolated quantum systems \cite{Wilczek2012}, subsequent theoretical studies have shown that periodically driven (Floquet) systems provide a suitable framework for their stabilization, leading to experimental observations in various platforms (see \cite{Zaletel2023,Sacha_review} for comprehensive reviews on the topic). Since then, the field has grown rapidly, with numerous advances in our understanding of their properties.

In addition to their theoretical importance in understanding fundamental aspects of non-equilibrium systems, time crystals may also hold promise for practical applications. Research in this direction is still in its infancy, with a recent growing interest in their use for metrology protocols \cite{Iemini2024,yousefjaniDiscreteTimeCrystal2024,yousefjaniDiscreteTimeCrystal2025,biswasFloquetCentralSpin2025,shuklaPrethermalFloquetTime2025,moonDiscreteTimeCrystal2024,gribbenQuantumEnhancementsEntropic2024,montenegroQuantumMetrologyBoundary2023a,pavlovQuantum2023,cabotContinuous2024,arumugamStarkmodulatedRydbergDissipative2025,liExactSteadyState2024,cabotQuantum2025}. However, the first steps have also been taken in the field of clocks \cite{singhQuantumThermodynamicsLimit2025,viottiQuantum2025}, complex network simulations \cite{estarellasSimulating2020}, quantum computation \cite{Bomantara2018}, thermal engines \cite{Carollo2020} and energy storage \cite{paulinoThermodynamics2024}. Notably, their use as quantum sensors has proven to be a particularly promising avenue, leveraging their unique properties in order to enhance the sensitivity and precision of measurement protocols. Recent works have analyzed their operation in several different models: disordered chains \cite{Iemini2024}, disorder-free ones with a gradient field \cite{yousefjaniDiscreteTimeCrystal2024,yousefjaniDiscreteTimeCrystal2025}, central spin models \cite{biswasFloquetCentralSpin2025}, prethermal platforms (multiferroic chains \cite{shuklaPrethermalFloquetTime2025, moonDiscreteTimeCrystal2024} or nitrogen-vacancy centers~\cite{rondin_magnetometry_2014}) or in dissipative systems \cite{gribbenQuantumEnhancementsEntropic2024,montenegroQuantumMetrologyBoundary2023a,pavlovQuantum2023,cabotContinuous2024,arumugamStarkmodulatedRydbergDissipative2025,liExactSteadyState2024,cabotQuantum2025}. Analyzing their dynamic operation is usually complicated. It is based on numerical simulations or relying on particular "fine-tuning" points in the specific model that allow for an analytical treatment. Understanding of their operation in general FTCs remains an open question.

\begin{figure*}
	\includegraphics[width=0.31\textwidth]{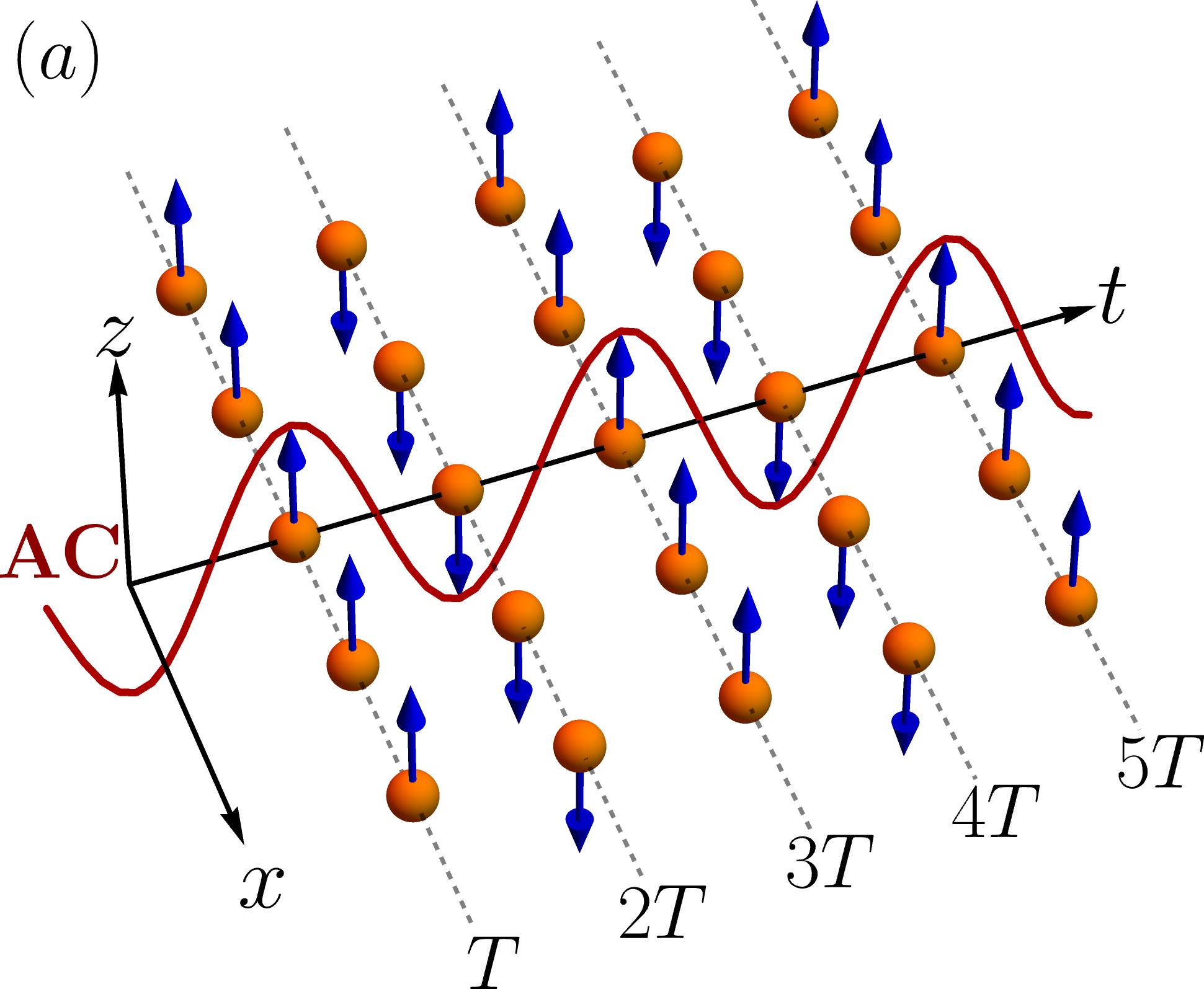}
	\includegraphics[width=0.25\textwidth]{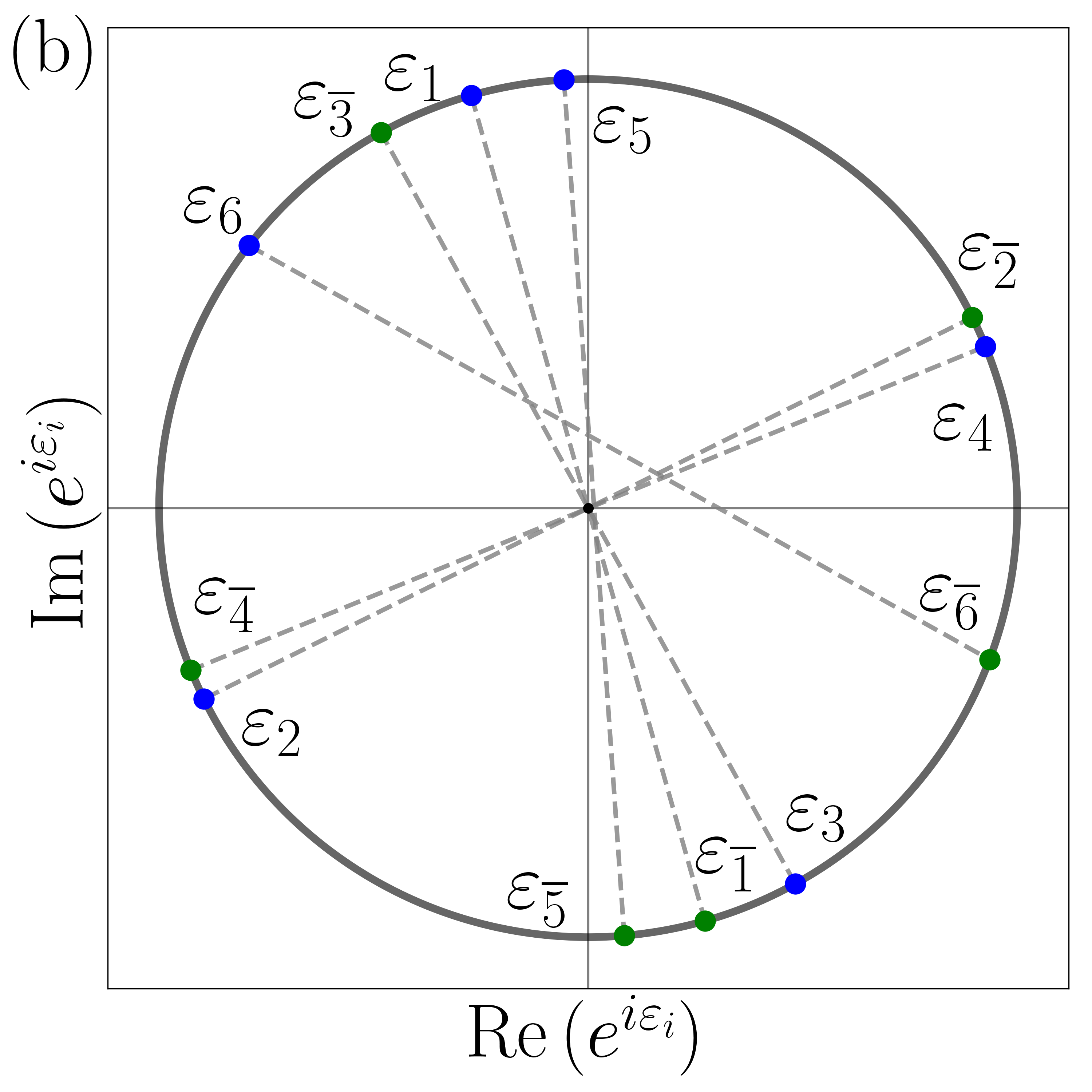}
	\includegraphics[width=0.41\textwidth]{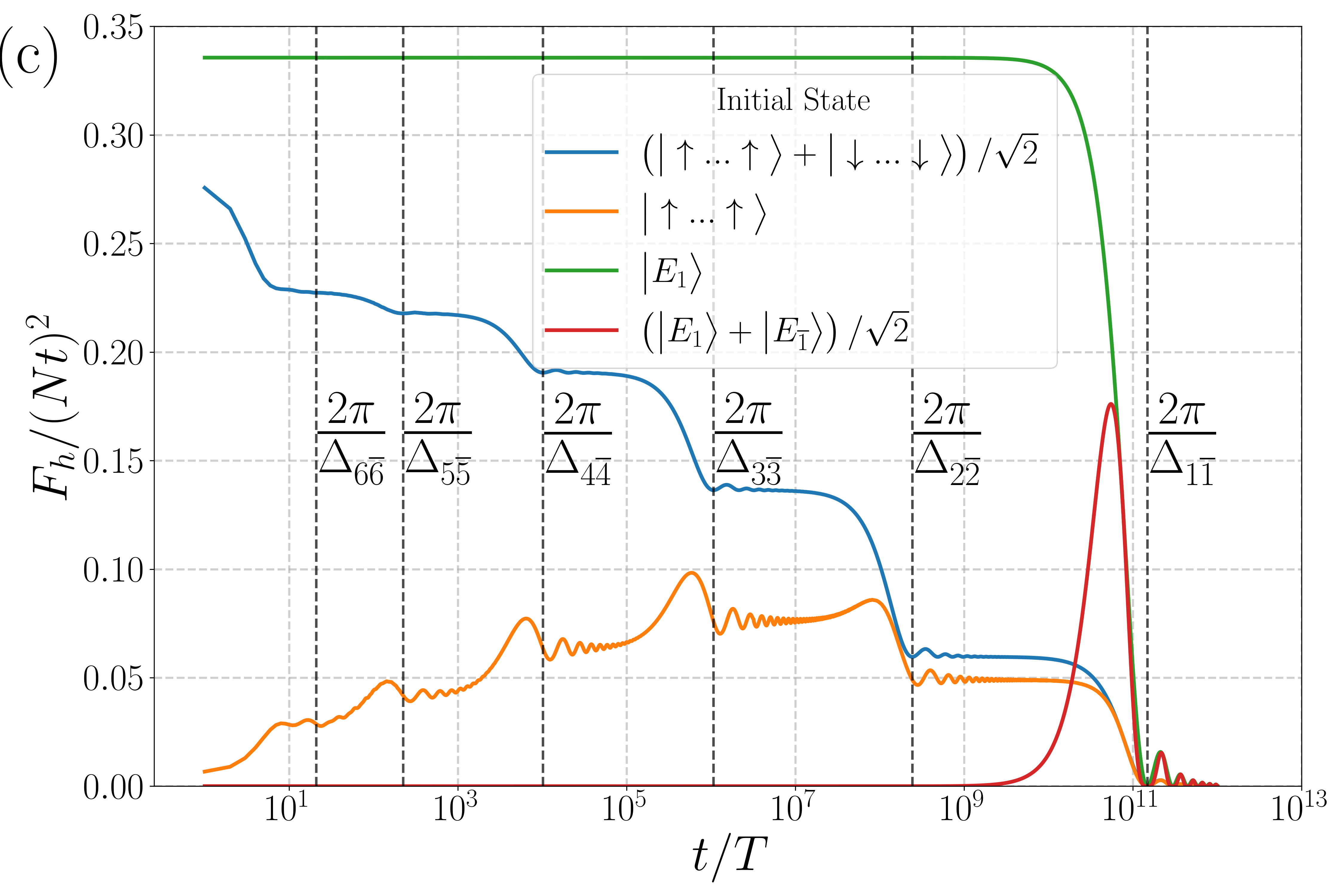}
	\caption{(a) Schematic representation of an FTC quantum sensor. The sensor consists of \(N\) spins and is exposed to an external AC field that acts as a probe for its amplitude or frequency.
	(b) Spectrum \(\{\epsilon_i\}\) of the Floquet unitary in the FTC sensor, showing the set of six $\pi$-paired quasienergies $(M=6)$ corresponding to cat states of opposite parities - the rest of the unpaired spectrum is omitted here for clarity.
	(c) Dynamics of the QFI for different relevant initial preparations of the sensor: from low to high correlated initial states, and effective Floquet Hamiltonian eigenstates \(\{ |E_{i(\bar{i})}\}\). The QFI exhibits a step-like increasing/decreasing characteristic dynamics for each of these classes of preparations, i.e.  a sequence of plateaus interspersed by abrupt variations, which occur
	on timescales proportional to their \(\pi\)-paired gaps (see Table~\ref{tab:states} for a more detailed summary).
	The results are illustrated here for an FTC sensor based on the LMG model~\cite{TsypilnikovLMGCode2025} (\(N=40\), \(T=1\), \(J=1\), \(B=0.4\), \(T_{\mathrm{AC}} = 2T\), linear response \(h \to 0\) and sinusoidal signal \(f(t)= \sin(\pi t/T)\)).
	See SM~\cite{SM}, Sec.~\ref{sec:sm-qfi-lmg-dynamics} for the extended data.
	}
	\label{fig:qfi-dynamics-lmg}
\end{figure*}

In this manuscript, we advance this field by presenting a general theory of Floquet time crystals (FTC) — including prethermal realizations — acting as AC sensors in closed systems. We elucidate their characteristic behavior through the analytical calculation of their optimal performance along the dynamics, i.e. the calculation of their quantum Fisher information (QFI).
The analytical results
(i) shed light on the mechanism behind the sensor operation; (ii) unravel a characteristic, step-like QFI dynamics; (iii) show the overall effects of approaching a phase transition; and (iv) demonstrate how exploring the direction and frequency of the AC field can achieve Heisenberg-limited precision for exponentially long times. See Fig.~\ref{fig:qfi-dynamics-lmg} for a schematic illustration. Thus, our results provide a solid theoretical basis for using such phases as sensors and pave the way to explore them — in their most diverse forms and possible experimental platforms — in metrological protocols, with potential applications in different fields.

The manuscript is organized as follows: in Sec.~\ref{sec:ftc}, we undertake FTC phases, discussing their main features and spectral structure. Sec.~\ref{sec:sensing-protocol} presents an overview of the metrology protocol, detailing the estimation precision bounded by the quantum Fisher information. We discuss our main results in Sec.~\ref{sec:ftc-sensors}, including analytical expressions for the QFI, along with a detailed discussion of their implications. To illustrate these findings, in Sec.~\ref{sec:lmg} we apply our framework to a specific FTC sensor based on the Lipkin-Meshkov-Glick (LMG) model. Finally, Sec.~\ref{sec:conclusions} concludes the work and provides perspectives on future research directions.

\section{Floquet time crystals}
\label{sec:ftc}
We consider a sensor system composed of \(N\) spins following a Floquet dynamics driven by \(\hat H_s(t) = \hat H_s(t+T)\), with \(T\) denoting the period. The Floquet unitary that describes the stroboscopic dynamics is given by \(\hat U_F = \mathcal{T}e^{-i\int_0^T \hat H_s(t')dt'}\), with \(\mathcal{T}\) denoting the time ordering operator. The structure of  \(\hat U_F\) in a FTC is assumed to be \cite{Zaletel2023}, 
\begin{equation}
	\label{eq:floquet-unitary}
	\hat U_F = \hat X e^{-i\hat H_F T}
\end{equation}
here \(\hat H_F\) is the Floquet Hamiltonian, \(\hat X^m = \mathbb{I}\) and \([\hat X, \hat H_F] = 0 \). The Floquet Hamiltonian is not necessarily equal to the sensor Hamiltonian \(\hat H_s(t)\), arriving  through a Magnus expansion of the Floquet unitary \cite{Bukov04032015}. In this particular expansion, the operator \(\hat X\) (which we denote parity operator)  corresponds to the underlying \(\mathbb{Z}_m\)  symmetry of the Floquet Hamiltonian. The existence of this decomposition directly leads to the possibility of FTCs. We focus our discussion  on the simplest scenario of \(m=2\), though the subsequent analysis can be adapted to encompass the broader case.

\textit{Spectral properties and cat-states.-}
Due to the commutativity of the operators, the Floquet unitary shares the same eigenstates as the Floquet Hamiltonian and the parity operator. Specifically,
\begin{equation}
	\label{eq:hf-floquet-schrodinger}
	\hat H_F |E_i \rangle = E_i |E_i \rangle, \quad \hat X |E_i \rangle = p_i |E_i \rangle,
\end{equation}
with \(p_i = \pm 1\) the ``parity'' of the \(i\)'th state, and therefore,
\begin{equation}
	\label{eq:floquet-eigenvalues}
	\hat U_F |E_i \rangle  = e^{-i \varepsilon_i } |E_i \rangle,\qquad
	\varepsilon_i = E_i T + \frac{(1-p_i)}{2} \pi,
\end{equation}
that is, the spectrum of the Floquet eigenstates shifts by \(\pi\) depending on its parity. Given that the Floquet Hamiltonian has an ergodicity-breaking mechanism, as spontaneous symmetry breaking (SSB) generated by the underlying \(\mathbb{Z}_2\), it has pairs of quasi-degenerate eigenstates with opposed parities for finite system sizes. These states are represented by the set \(\{|E_i\rangle,| E_{\bar{i}}\rangle\}_{i=1}^M\) with opposed parities, where \(M\leq d_H\) with \(d_H\) the Hilbert space dimension and the index \((i,\bar{i})\) represents the \(i\)'th paired states. The gap \(\Delta_{i\bar{i}} = E_i - E_{\bar{i}} \) decays with system size (e.g. exponentially \(\mathcal{O}(e^{-N})\)), becoming fully degenerate in the macroscopic limit.
The corresponding Floquet eigenvalues are thus \(\pi\)-paired, with \(\varepsilon_i - \varepsilon_{\bar{i}} = \pi - \Delta_{i\bar{i}} T\) (see Fig.~\ref{fig:qfi-dynamics-lmg}(b)). Moreover, due to the symmetry of the system, these eigenstates are in the form of cat states,
\begin{equation}
	\label{eq:cat-states}
	|E_i\rangle \propto |\Uparrow_i\rangle + p_i |\Downarrow_i\rangle,
\end{equation}
where \(\hat X |\Uparrow_i (\Downarrow_i)\rangle = |\Downarrow_i (\Uparrow_i) \rangle\), and \(|\Uparrow_i(\Downarrow_i)\rangle \) are macroscopically distinct states representing the decomposition of the cat state in the broken symmetry sectors (see SM~\cite{SM} Sec.~\ref{sec:sm-ssb-cat} for a discussion on the detailed structure of the cat states).

An observable able to capture the FTC phenomenology is given by a Hermitian operator \(\hat S_z\) anti-commuting with the symmetry operator,
\begin{equation}
	\label{eq:anticomm-x-sz}
	\{ \hat S_z, \hat X \} = 0, \qquad \hat X \hat S_z \hat X = -\hat S_z,
\end{equation}
where,
\begin{equation}
	\label{eq:sz-matrix-elements}
	\langle E_i | \hat S_z |E_j \rangle =  \langle E_{\bar{i}} | \hat S_z |  E_{\bar{j}} \rangle = 0,
\end{equation}
and the only nonzero elements of the operator are those among different parities. Therefore, one obtains the dynamics,

\begin{eqnarray}
	\label{eq:sz-expand}
	\hat S_z(nT) &=& (\hat U_F^\dagger)^n \hat S_z (\hat U_F)^n \\
	&=& (-1)^{n} \sum_{i,j} e^{i\Delta_{ij}nT} \langle E_i | \hat S_z |E_j \rangle |E_i\rangle \langle E_j|\nonumber
\end{eqnarray}

where we expanded \(\hat S_z\) in the Floquet unitary eigenbasis, and used the commutation relation among the operators.
The observable oscillates in a period doubling way (\((-1)^{n}\)) apart from dephasing terms (\(e^{\Delta_{ij} nT }\)) among different parity sectors. Despite the majority of these terms may vanish upon an average over time, those corresponding to \(\pi\)-paired cat states are stable due to the vanishing small gap. Therefore the system features a long lived period doubling oscillations lasting a time inversely proportional to their gap (\(\Delta_{i\bar{i}}^{-1}\)), breaking the discrete time symmetry and stabilizing a FTC in the macroscopic limit.

\section{Sensing Protocol}
\label{sec:sensing-protocol}

The proposed sensing scheme is illustrated in Fig.~\ref{fig:qfi-dynamics-lmg}(a).
A sensor in the FTC phase is put in contact with an external periodic field \(\hat V(t)=h f(t) \hat O\),  where \(h\) is its amplitude, \(f(t)=f(t+T_{\mathrm{AC}})\) the time modulation with \(T_{\mathrm{AC}}\) its periodicity, and \(\hat O\) the direction of the field. The sensor coupled to the external field evolves under the Hamiltonian \(\hat H(t) = \hat H_s(t)+\hat V(t)\).
The goal of the sensor is to estimate an unknown parameter, which in our case is the amplitude \(h\) of the field (although the reasoning of our analysis could be extended to other parameters of the external field).

In the frequentist framework for sensing, the standard estimation protocol follows a repeating sequence of steps: (i) initializing the sensor in an advantageous (possibly entangled) state; (ii) allowing the sensor to interact with the signal of interest (\(h\)) for a set duration, imprinting the unknown parameter onto the sensor's state; and (iii) performing a measurement on the quantum sensor. By accumulating statistics from repeated trials, the parameter \(h\) can be estimated with optimal precision.
If prior knowledge about the parameter is available, the estimation uncertainty \(\Delta h(t)\) is bounded by the quantum Cramér-Rao bound,
\begin{equation}
	\label{eq:cramer-rao}
	\Delta h(t) \geq \frac{1}{\sqrt{\mu F_h(t)}}
\end{equation}
where \(F_h(t)\) represents the quantum Fisher information (QFI) of the probe and \(\mu\) is the number of measurements~\cite{Braunstein1994,Liu2019}. The QFI has several key properties worth noting. In classical systems, the QFI is constrained by the number of repetitions in the estimation protocol, leading to a linear scaling in time. However, leveraging quantum coherence enables a quadratic enhancement in time, extending the QFI’s maximum to
\(F_h(t)\leq t^2\)~\cite{Pang2017}.

Additionally, in  many-body systems one can consider another scaling to the QFI related to the number of particles participating in the sensor. Since the Fisher information is additive, a sensor comprising \(N\) separable particles adheres to the bound \(F_h(t) \leq N t^2\), also called the standard quantum limit (SQL). Surpassing this limit requires nonseparability, therefore exploiting  in a beneficial form the quantum correlations among the particles. The ultimate quantum advantage arises when the quantum correlations are fully utilized, yielding a quadratic improvement in both particle number and time. In that case, this achieves the Heisenberg limit, where \(F_h(t) = N^2 t^2\).

For a pure state the QFI has the form~\cite{Iemini2024},
\begin{equation}
	\label{eq:qfi-pure}
	\frac{F_h(t)}{4} =  \langle \psi(0)|\hat S^\dagger_h(t) S_h(t) |\psi(0)\rangle - \left|\langle \psi(0)|\hat S_h(t) |\psi(0)\rangle \right|^2
\end{equation}
where \(\hat S_h(t)\) characterizes the evolved signal operator, which we denote as the ``Heisenberg signal operator'' (HSO),
\begin{equation}
	\label{eq:hso}
	\hat S_h(t) = \int_0^t \hat U^\dagger_h(t') \left(
	\frac{\partial \hat H(t')}{ \partial h} \right)
	\hat U_h(t') dt'
\end{equation}
with \(\hat U_h(t) = \mathcal{T} e^{-i \int_{0}^{t} \hat H_s(t') + \hat V(t') dt'}\) the unitary operator for the dynamics.

\begin{figure*}
	\includegraphics[width=\textwidth]{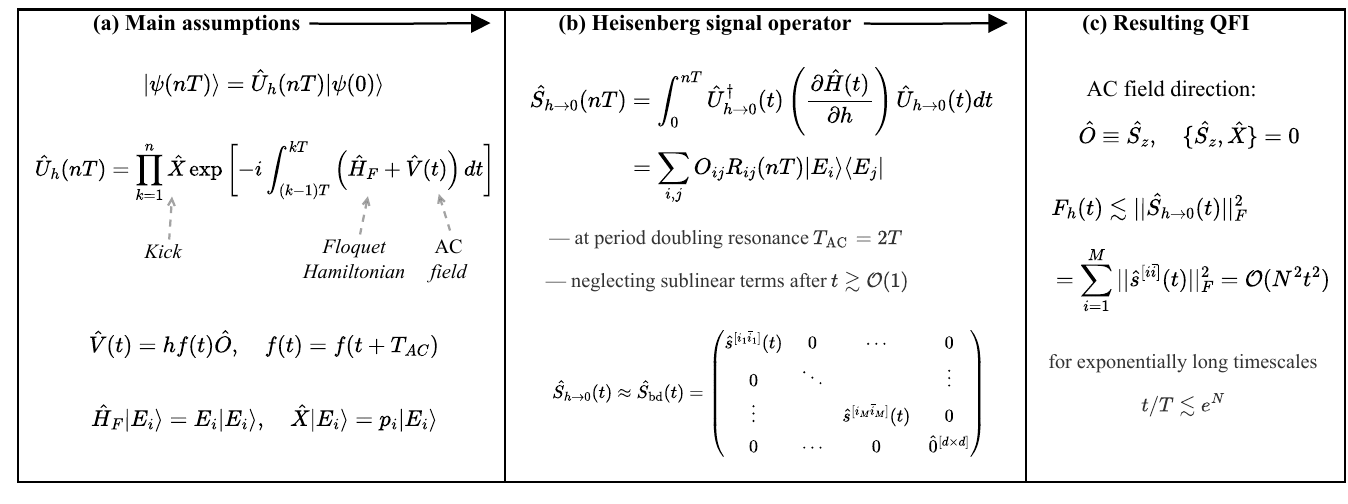}
	\caption{
		\textit{QFI derivation outline: } (a) The FTC sensor, governed by a Floquet Hamiltonian \(\hat H_F\) and parity operator \(\hat X\), is put in contact with an AC field \(\hat V(t)\). Without loss of generality, we assume that the dynamics are driven by the shifted Floquet Hamiltonian, as Eq.~\eqref{eq:hf-shift}. (b) Tuning the AC field frequency in PDR to the sensor simplifies the HSO, leading to a block-diagonal structure along the \(\pi\)-paired subspaces (Eq.~\eqref{eq:hso-block-diagonal}). (c) Aligning the AC field along the underlying SSB direction achieves an HSO norm (which upper-bounds the QFI) sustaining the Heisenberg limit scaling for exponentially long timescales.
	}
	\label{fig:derivation-sketch}
\end{figure*}

\section{FTC Sensors}
\label{sec:ftc-sensors}

This section analyzes the QFI and HSO of the FTC sensor. We begin with an intuitive discussion of the sensor operation,  revealing how the period-doubling resonance (PDR) mechanism can potentially lead to long-lived Heisenberg scaling. Next, we derive analytically the HSO and obtain the characteristic QFI dynamics for different state preparations, including parity eigenstates, symmetry-breaking states, and states confined to a single cat subspace. We subsequently analyze the behavior across the phase transition and, finally, examine how these results evolve in the nonlinear-response regime.

A simplistic, but intuitive picture, of the sensor's performance can be drawn as follows. Assume that the FTC is in its stable period doubling regime, and lies in a cat subspace whose states are (apart from their parities) roughly equal, \(|\Uparrow_i(\Downarrow_i)\rangle \approx
	|\Uparrow_{\bar{i}}(\Downarrow_{\bar{i}})\rangle\). The dynamics of the FTC with no AC field is approximated by,
\begin{equation}
	\label{eq:psi-intuitive-dynamics}
	|\psi(t)\rangle = c_i(t)|E_i\rangle + c_{\bar{i}}(t)|E_{\bar{i}}\rangle \approx
	c_\uparrow(t)|\Uparrow_i \rangle + c_{\downarrow}(t) |\Downarrow_i \rangle
\end{equation}
with \(c_{\uparrow(\downarrow)}(t+T) = c_{\downarrow(\uparrow)}(t)\). If we add on top of it an external AC field lying along their \(\Uparrow_i(\Downarrow_i)\) direction, such that \({e^{i f(t) h \hat O} |\Uparrow_i(\Downarrow_i)\rangle} = e^{\pm i f(t) h} |\Uparrow_i(\Downarrow_i)\rangle\) the sensor will accumulate a local phase for a stroboscopic period  \( \Theta(t) = e^{-2 i h \int_t^{t+T} f(t')dt'}\) . Tuning the AC frequency  in PDR resonance to the internal FTC spin dynamics (i.e. \(f(t+T)=-f(t)\), \(|f(t)|=1\)) the phase is accumulated constructively along all dynamics,
\begin{equation}
	\label{eq:psit-intuitive}
	|\psi(t)\rangle = c_\uparrow(t)|\Uparrow_i \rangle +  e^{- 2 i ht} c_{\downarrow}(t) |\Downarrow_i \rangle.
\end{equation}
In this way, the AC amplitude information is encoded coherently in the sensor, leading to a high precision for its estimation in the metrological protocol. This simple picture therefore indicates that the sensor will perform well over exponentially long FTC lifetimes. Since the information is encoded in cat states, it also shows the potential benefits of their entanglement in boosting the corresponding QFI to the Heisenberg limit.

The dynamics of the system, however, is generally more complex than such an ansatz description. Additionally, an analytical calculation of the QFI involves the entire Hilbert space spectrum and its susceptibility to the perturbed AC field, making it a nontrivial task.
We obtain here, however, a simplified analytical expression for the QFI elucidating some of its general behavior.

In our approach we assume that the action of the external AC field is a shift on the effective Floquet Hamiltonian of the FTC,
\begin{equation}
	\label{eq:hf-shift}
	\hat H_F \rightarrow \hat H_F + \hat V(t).
\end{equation}
This assumption corresponds, for example, to the case where the FTC Hamiltonian is already in its  Floquet Hamiltonian form,
\begin{equation}
	\label{eq:floquet-hamiltonian-form}
	\hat H_s(t) = \hat H_F +  \sum_{m=1}^\infty \delta(t-mT) \ln (\hat X)
\end{equation}
Here, it appears to be a restricted group of FTC sensors. However, we expect this same phenomenology to occur also for general FTC Hamiltonians (not necessarily in the above form) by reasoning with a simpler — but not simplistic — form of AC signals. Specifically, point-like stroboscopic signals
\begin{equation}
	\label{eq:f-pointlike}
	f(t) = \sum_{m=1}^\infty \delta(t-mT) g(t), \quad g(t) = g(t+T_{\mathrm{AC}}),
\end{equation}
despite their simplicity, still carry the key characteristics of any modulated signal --
namely, a definite frequency and a given strength/amplitude. Therefore, we expect them to have the same effect on the sensor as more general signals with the same strength and frequency, but more intricate envelope structures. In fact, the above example (Eq.~\eqref{eq:psit-intuitive}) intuitively shows that the sensor performance relies on the accumulated local phase \(\Theta(t)\) along the stroboscopic dynamics. Therefore, the integrated signal over the AC period for different envelopes will lead to qualitatively different factors, but similar time-dependent behaviors.

Importantly, in addition to this qualitative equivalence between the signals, note that if the point-like stroboscopic signal is a subharmonic of the sensor periodicity (\(T_{\mathrm{AC}} = nT, n\geq 1\)), then due to its point-like nature, the relation of Eq.~\eqref{eq:hf-shift} is valid for any FTC Hamiltonian, \(\hat H_s(t)\). Therefore, despite the assumption Eq.~\eqref{eq:hf-shift} appearing restrictive, we expect the results to be qualitatively similar to it in the general case.

\subsection{Linear response (LR)}
Let us first discuss the case where the amplitude is negligible as compared to any other parameter in the system (e.g. time, energy), denoted as linear response \(h\rightarrow 0\). A schematic summary of the derivation is shown in Fig.~\ref{fig:derivation-sketch}. We first obtain that the HSO can be expanded as follows (details of the derivation in SM~\cite{SM} Sec.~\ref{sec:sm-derivation-qfi}),
\begin{equation}
	\label{eq:qfi-linear-response}
	\hat S_{h\rightarrow 0}(nT) =\sum_{i,j}  O_{ij}  R_{ij}(nT) |E_i\rangle \langle E_j|,
\end{equation}
where \( O_{ij} = \langle E_i| \hat O | E_j\rangle\) and
\( R_{ij}(nT) =  \sum_{k=0}^{n-1} g_{k} (p_i p_j)^k e^{i k \Delta_{ij} T }\) is a weighted geometric progression, with
\begin{equation}
	\label{eq:pk-linear-response}
	g_{k} =  \int_{0}^{T} f(kT+t')   e^{i \Delta_{ij} t' }dt'
\end{equation}
The expansion can be interpreted with the overlapping terms \(\hat O_{ij}\) representing the  potential of the AC field operator to correlate the different eigenstates in the dynamics, with the \(\hat R\) matrix their corresponding time-dependent response to the applied signal. In order to amplify the QFI one intuitively aims to maximize these terms, since  \(F_h(t) \leq || \hat S_h(t)||_F^2\)  with \(||\hat S_h(t) ||_F=\sqrt{Tr (\hat S_h(t)^\dagger \hat S_h(t))}\) the Frobenius operator norm. Considering an AC operator along the SSB direction of the underlying cat-states, one can amplify the corresponding  overlapping elements \( O_{i\bar{i}} \equiv (\hat S_z)_{i\bar{i}} \sim \mathcal{O}(N)\), scaling with the number of spins in the sensor. The periodic modulation \((p_i p_j)^k\) in the geometric   progression indicates a natural PDR resonance frequency for the sensor, \(f(t+T) = -f(t)\).
Moreover, under PDR the response terms \(R_{ij}(t)\) dephase for times larger than their inverse gap (\(t  \gtrsim t_{ij}^* \equiv \Delta^{-1}_{ij}\)). Therefore, after an initial transient time of the order \(t \sim \mathcal{O}(\Delta_{i,j\neq i,\bar{i}}^{-1})\) the HSO reduces to a block diagonal form along \(\pi\)-paired cat subspaces (details of the derivation in SM~\cite{SM} Sec.~\ref{sec:sm-derivation-qfi}),

\begin{equation}
	\label{eq:hso-block-diagonal}
	\hat{S}_{h \rightarrow 0}(t) \approx  \hat{S}_{\mathrm{bd}}(t) =
	\begin{pmatrix}
		\hat{s}^{[i_1\bar{i}_1]}(t) & 0      & \hdots                      & 0                      \\
		0                           & \ddots &                             & \vdots                 \\
		\vdots                      &        & \hat{s}^{[i_M\bar{i}_M]}(t) & 0                      \\
		0                           & \hdots & 0                           & \hat{0}^{[d \times d]}
	\end{pmatrix}
\end{equation}
Here, the approximation neglects
terms that are sublinear in time, and \(\hat s^{[i \bar{i}]}(t) \equiv (\hat S_{h \to 0}(t))_{m,n=(i,\bar{i})}\) are block-diagonal terms. Specifically, these are \(2 \times 2\) matrices, whose elements are those of the paired cat-states:
\begin{equation*}
	\hat{s}^{[i \bar{i}]}(t) =
	\begin{pmatrix}
		O_{ii}R_{i i}(t)             & O_{i \bar{i}} R_{i \bar{i}}(t)            \\
		O_{\bar{i} i}R_{\bar{i}i}(t) & O_{\bar{i} \bar{i}}R_{\bar{i} \bar{i}}(t)
	\end{pmatrix},
\end{equation*}
and \(\hat{0}^{[d \times d]}\) represents the null matrix of size \(d \times d\), with \(d = d_H - 2M\) corresponding to the dimension of the remaining spectrum. The HSO norm thus saturates the Heisenberg limit for exponentially long times, \(||\hat S_{bd}(t/T \lesssim e^{N})||_F^2 = \mathcal{O}(N^2t^2) \). This can be seen from the fact that (i) the squared norm of a block-diagonal matrix is equal to the sum of the norms of its blocks, \(||\hat S_{bd}(t)||_F^2 = \sum_{i=1}^M ||\hat s^{[i \bar{i}]}(t)||_F^2\), and (ii) for times smaller than the inverse gaps (\(t_{i\bar{i}}^*\)), which are exponentially long with the system size, each of these blocks has a norm that scales linearly with time and the number of spins, \(||\hat s^{[i \bar{i}]}(t)||_F = \mathcal{O}(Nt)\).

The HSO can equivalently be interpreted as generating the leading order corrections in the perturbed sensor dynamics. The emergence of the block diagonal structure highlights the mechanism behind the sensor operation,  that is triggering resonant transitions between pairs of cat states during the dynamics.
Note also that the sensor operation is based on the underlying many-body FTC phase, making the protocol robust. Perturbations in the sensor Hamiltonian are translated into corrections in the \(\pi\)-pairing states (and gaps) as well as in the emergent SSB direction (i.e., in the corresponding \(\hat S_z\) operator). However, these corrections will only affect the sensor at its qualitative level, shifting the sensor lifetime proportional to the FTC gaps — and in the overlapping terms, which, although the AC field being no longer completely aligned with the emergent SSB direction, would still have a large part along this preferred direction.\\

\begin{table*}[t]
	\centering
	\begin{tabular}{|c|c|c|}
		\hline
		\multicolumn{3}{|c|}{
		\rule{0pt}{5ex} 
		Characteristic step-like dynamics:
		\(\quad \delta_k F_h \equiv \dfrac{F_h(t \gtrsim \Delta^{-1}_{k \bar{k}}) - F_h(t \lesssim \Delta^{-1}_{k \bar{k}} )}{t^2}\)
		\rule[-2ex]{0pt}{0pt} 
		}                                                                                                                                                                                                                                                                                                                                                                           \\
		\hline 
		\multicolumn{2}{|c|}{Initial State} & Behavior                                                                                                                                                                                                                                                                                                                              \\
		\hline \rule{0pt}{5ex} 
		Parity initial state                & \(\sum_{i=1}^{M} \sqrt{c_i} |E_i \rangle +  |\chi_{\perp,\rm{cat}} \rangle\)        & \(\delta_k F_h \propto -c_k |O_{k\bar{k}}|^2 <0\), strictly decreasing steps  \rule[-3ex]{0pt}{0pt}                                                                                                                                                 \\
		\hline \rule{0pt}{5ex} 
		SSB state                           & \(\sum_{i=1}^{M} \sqrt{c_i} |\Uparrow_i \rangle +  |\chi_{\perp,\rm{cat}} \rangle\) & \(\delta_k F_h \propto O_{k\bar{k}} c_k (\mathcal{P}_{M_k} - \mathcal{P}_{\perp M_k})\), increasing and decreasing steps \rule[-3ex]{0pt}{0pt}                                                                                                      \\
		\hline \rule{0pt}{5ex} 
		Single cat subspace state           & \(\cos(\theta/2) |E_k\rangle + e^{i\varphi }\sin(\theta/2) |E_{\bar{k}}\rangle\)    & \(\frac{F_h(t)}{t^2} \propto \left(1-\cos^2{ \left(t \Delta_{k\bar{k}} / 2 + \varphi \right)}\sin^2{\theta}  \right) \frac{\sin^2{\left(t \Delta_{k\bar{k}}/2\right)}}{\left(t \Delta_{k\bar{k}} / 2\right)^2}\), single step \rule[-3ex]{0pt}{0pt} \\
		\hline
	\end{tabular}
	\caption{Summary of the step-like QFI dynamics for different initial state preparations, highlighting cases with fixed parity, symmetry-broken, and single-subspace cat states.}
	\label{tab:states}
\end{table*}

\textbf{Characteristic step-like dynamics:} Under PDR the block diagonal terms dephase sublinearly in time at \(\{ t^*_{i\bar{i}}\} \). Therefore, the QFI renormalized by the SQL, \(F_h(t)/Nt^2\), will generally exhibit a step-like behavior on these time scales, as illustrated in Fig.~\ref{fig:qfi-dynamics-lmg}(c) and summarized in Table~\ref{tab:states} for different initial state preparations. The step difference due to the k'th block sector is given by,
\begin{equation}
	\label{eq:step-qfi-definition}
	\delta_k F_h \equiv \frac{F_h(t \gtrsim t^*_{k \bar{k}}) - F_h(t \lesssim t^*_{k \bar{k}} )}{t^2}.
\end{equation}
These are determined by the initial preparation of the sensor and the AC magnetization before the corresponding characteristic time.
Defining the projector in the cat subspace, \(\hat P_{\rm cat} = \sum_{i=1}^M |E_i\rangle \langle E_i|\), the projection of the initial state on such subspace dictates the long-time QFI dynamics.

\textit{ (i) Parity initial states.- } Initial states with fixed parity have a particularly simple step-like QFI dynamics. We consider,

\begin{equation}
	\label{eq:init-parity}
	|\psi(0) \rangle = \sum_{i=1}^{M} \sqrt{c_i} |E_i \rangle +  |\chi_{\perp,\rm{cat}} \rangle
\end{equation}
where \(c_i \ge 0\), the first term represents the \(M\) cat eigenstates with fixed parity  \(p_\psi\), \(\hat X |E_i \rangle = p_\psi |E_i \rangle\), and \(|\chi_{\perp,\rm{cat}} \rangle = (\mathbb{I}-\hat P_{\rm cat})|\psi(0)\rangle \) the portion out of it.
Due to the parity symmetry, the QFI after the initial transient time  is given by \(F_h(t)/t^2 = \sum_{i=1}^M  c_i |O_{i\bar{i}}|^2\). Assuming the expansion coefficients \(\{c_i\}\) have no strong dependence with \(N\), the QFI thus scales quadratically with the number of spins, since \(|O_{i\bar{i}}|=\mathcal{O}(N)\).
Furthermore, \(\delta_k F_h =-c_k |O_{k\bar{k}}|^2 <0,\,  \forall k\), the steps are always decreasing in time.
The general behavior of the QFI thus follows a Heisenberg limit scaling (\(F_h(t) \sim N^2 t^2\)) arising from the initial preparation in high correlated states, followed by a sequence of decreasing steps due to dephasing in each of the subspaces, until the sensor performance is completely degraded.

\textit{ (ii) Symmetry breaking states.- } Initial states breaking  the Floquet Hamiltonian symmetry, such as spins aligned along the \(\hat S_z\) direction, usually have low correlations and are therefore easier to prepare in an experimental setting. We consider the class of initial states,

\begin{equation}
	\label{eq:init-ssb}
	|\psi(0) \rangle = \sum_{i=1}^{M} \sqrt{c_i} |\Uparrow_i \rangle +  |\chi_{\perp,\rm{cat}} \rangle
\end{equation}
where \(c_i  \ge 0 \) and the first term explicitly breaks the symmetry.
For sufficiently large system sizes pairs of broken symmetry states are similar (\(|\Uparrow_i (\Downarrow_i) \rangle \sim
	|\Uparrow_{\bar{i}} (\Downarrow_{\bar{i}}) \rangle\)) and
the step difference reduces to (see SM~\cite{SM} Sec.~\ref{sec:sm-symmetry-breaking}),

\begin{equation}
	\label{eq:step-qfi-ssb}
	\delta_k F_h \propto O_{k\bar{k}} c_k (\mathcal{P}_{M_k} - \mathcal{P}_{\perp M_k})
\end{equation}
where \(\mathcal{P}_{M_k} = \sum_{i \in M_k} c_i\) is the overlap of the initial state within the \(M_k\) subspace,
\(\mathcal{P}_{\perp M_k} = \sum_{i \in M, i \notin M_k} c_i\) its overlap in the cat subspace, but out of \(M_k\), and \(M_k\) the set of cat state pairs with a gap smaller (or equal) than the k'th pair.
The QFI dynamics is thus characterized by a first sequence of increasing steps, while \(\mathcal{P}_{M_{k}} > \mathcal{P}_{\perp M_{k}}\). In this regime, while the first pairs of cat states are dephased there are large correlations being created among the remaining ones with longer lifetimes, which leads to increasing steps. Eventually, these also start to dephase leading to the further set of now decreasing steps in the sensor performance.

\textit{  (iii) Initial states in a single cat subspace.- } in the case of e.g. an exact ground state or any combination with its closest energetic cat pair state due to lack of full resolution in the experimental apparatus, we derive the full analytical expression of the QFI dynamics. Specifically, given
\begin{equation}
	|\psi_{i\bar{i}}(0)\rangle = \cos(\theta/2) |E_i\rangle + e^{i\varphi }\sin(\theta/2) |E_{\bar{i}}\rangle
\end{equation}
and considering explicitly a sinusoidal signal,
\(f(t) = \sin(\pi t/T + \phi_{\mathrm{AC}})\), we obtain that

\begin{eqnarray}
	\label{eq:qfi-single-cat}
	\frac{F_h(t)}{|O_{i\bar{i}}|^2 t^2} &= & \frac{16 \cos^2 { \phi_{\mathrm{AC}}} }{\pi^2} \left(1-\cos^2{ \left(\frac{ \Delta_{i\bar{i}} t}{2} + \varphi \right)}\sin^2{\theta}  \right) \nonumber \\  & & \hspace{2.5cm} \times \left( \frac{ \sin(\Delta_{i\bar{i}}t/2) }{ (\Delta_{i\bar{i}}t/2)}\right)^2
\end{eqnarray}

We see that it recovers the previous behaviors at its extremum phases, i.e. \(\theta =0,\pi\) (Eq.~\eqref{eq:init-parity}) or \(\theta = \pi/2\) (Eq.~\eqref{eq:init-ssb}) with \(\phi_{\mathrm{AC}}=0\).
In the latter case, the optimum renormalized QFI (peaked value) occurs at \(t_{\rm max} \approx   2.331 \Delta_{i\bar{i}}^{-1}\), which gives  a maximum \(F_h(t_{\rm max})/|O_{i\bar{i}}|^2 t^2\approx  0.523\frac{16 \cos^2 \phi_{\mathrm{AC}}}{\pi^2}\), as shown in SM~\cite{SM} Sec.~\ref{sec:sm-single-subspace}. Interestingly, despite its small initial QFI and low correlations, if the sensor is operated for sufficiently long times it can still achieve a roughly equivalent scaling to that maintained by highly correlated initial preparations. Most of the state correlations generated by the FTC are naturally absorbed in the sensing protocol.

\subsection{QFI along quantum phase transitions} Given the Floquet Hamiltonian has a phase transition at a critical coupling \(g_c\), its order parameter has a discontinuity as \(|O_{i \bar{i}}|/N \sim |g-g_c|^z\) with critical exponent \(z\). Therefore, according to our previous results, after the initial transient time the QFI scales as 
\begin{equation}
	\label{eq:qfi-critical-scaling}
	F_h(t) \sim N^2 t^2 |g-g_c|^z,
\end{equation}
apart from its step-like dynamics due to initial preparation. In contrast to critical sensing, which exploits the high susceptibility of a criticality to enhance the sensor operation, here we see no direct amplification; rather, a qualitative decreasing on top of the Heisenberg limit as one approaches the critical coupling.
Worth remarking that exactly at the critical point, the gap between different cat states may vanish with system size. Given their vanishing is slower than that between cat pairs, \(\lim_{N \rightarrow \infty} \Delta_{ij}/\Delta_{i\bar{i}} = 0\),  \(\forall i,j \neq \bar{i}\), our conclusions are maintained due to the emergence of the block diagonal structure for the HSO (Eq.~\eqref{eq:hso-block-diagonal}). In the opposite case, however, there is no clear separation of timescales and our results do not hold, requiring a more careful analysis.


\subsection{Nonlinear response (NLR)}
We extend our analysis to larger AC amplitude fields, by considering a Heaviside signal in PDR to the sensor (\(f(t+T) = -f(t)\) with \(|f(t)|=1\)) manageable for an analytical treatment.
We find that the HSO has the same structure as Eq.~\eqref{eq:qfi-linear-response} with the modified eigendecomposition,
\begin{equation}
	\label{eq:nlr-shift-notation}
	|E_i \rangle \rightarrow  |E_{i,h} \rangle,\quad
	\Delta_{ij} \rightarrow  \Delta_{ij,h} = E_{i,h} - E_{j,h},
\end{equation}
(and corresponding \( O_{ij,h}\), \(R(t)_{ij,h}\) terms) where \(\{ E_{i,h}, |E_{i,h}\rangle \}\) are the eigenvalues and eigenvectors of the effective Floquet Hamiltonian with a constant symmetry breaking perturbation, \(\hat H_{F,h}^{[\rm effective]} = \hat H_F + h \hat S_z\) (see SM~\cite{SM} Sec.~\ref{sec:sm-effective-floquet}). The AC field thus opens \(\pi\)-pairing gaps further accelerating the resonant transitions and the subsequent dephasing on such subspaces. It induces a new characteristic time scale \(t_{\rm NLR,i}\): while for \(t \lesssim t_{\rm NLR,i}\) the sensor behaves as in LR, for longer times \(t \gtrsim  t_{\rm NLR,i}\) the QFI features a constant quadratic scaling in time. Given that the AC field weakly perturbs the spectrum,
\begin{equation}
	\label{eq:pt-condition}
	h |O_{i\bar{i},0}| \ll E_i - E_{j}, \quad  \forall i \mbox{ and } j \neq \bar{i}.
\end{equation}
using perturbation theory (and assuming for simplicity that \(O_{i\bar{i}} \in \mathbb{R}\)) we obtain that (see derivation in SM~\cite{SM} Sec.~\ref{sec:sm-nonlinear-response})
\begin{equation}
	\label{eq:tnlr-time}
	t_{\rm{NLR},i} = \left( 2h |O_{i\bar{i},0}| \sqrt{1+ \left(\frac{2h |O_{i\bar{i},0}|}{\Delta_{i\bar{i},0} }\right)^2 } \right)^{-1}
\end{equation}
We see that, apart from corrections of the order of \(2h |O_{i\bar{i},0}|/\Delta_{i\bar{i},0}\), the NLR effects settle into the dynamics at times inversely proportional to the perturbation, \(t\sim 1/2h |O_{i\bar{i},0}|\).

\section{LMG model}
\label{sec:lmg}

An FTC sensor based on the Lipkin-Meshkov-Glick (LMG) Hamiltonian~\cite{Lipkin1965, russomanno_floquet_2017} provides a suitable illustration of the theory discussed above. The sensor is composed of \(N\) spins with infinite-range interaction
\begin{eqnarray}
	\label{eq:lmg-hs-full}
	\hat{H}_s(t) &=& \hat{H}_{\text{LMG}} + \hat{H}_{\text{Kick}}(t) + \hat V(t),
\end{eqnarray}
with
\begin{eqnarray}\label{eq:lmg-hamiltonian}
	\hat{H}_{\text{LMG}} &=& -\frac{2J}{N} \hat{S}_z^2 - 2 B  \hat{S}_x, \nonumber \\
	\hat{H}_{\text{Kick}}(t) &=& -\pi \sum_{n=1}^{\infty} \delta(t - nT) \hat{S}_x, \\
	\hat{V}(t) &=& h f(t) \hat{S}_z, \nonumber
\end{eqnarray}
where \(\hat S_\alpha = \sum_{i=1}^N \hat \sigma_i^\alpha/2\), \(\alpha = x,y,z\) are collective spin operators, with  \(\hat \sigma_i^\alpha\) Pauli operators acting on the i'th spin. The term \(J\) denotes the exchange interaction associated with infinite-range coupling, with a prefactor \(1/N\) ensuring a finite free energy per spin in the macroscopic  limit, and \(B\) is the strength of a uniform magnetic field along the transverse \(x\)-direction. The kick occurs at stroboscopic times inducing a \(\pi\) rotation around the \(x\)-direction. Notice that without AC field, the sensor Hamiltonian is already in the Floquet unitary structure as discussed in Eq.~\eqref{eq:floquet-unitary}, with the LMG Hamiltonian playing the role of the Floquet Hamiltonian (\(\hat H_{\rm LMG}=\hat H_{F} \)), and the kick term as its  \(\mathbb{Z}_2\) parity symmetry (\(e^{-i\hat H_{\rm Kick}} = \hat X = \prod_{i=1}^N \hat \sigma_i^x\)), where \([\hat X, \hat H_{\rm LMG}] = 0\).
The AC field of amplitude \(h\) is modulated by a sinusoidal signal in PDR to the FTC, \(f(t) = \sin(\pi t/T)\), and lying along the underlying SSB direction (\(\hat O \equiv \hat S_z\) , satisfying Eq.~\eqref{eq:anticomm-x-sz}, which corresponds exactly to the collective spin operator in the LMG model) in order to maximize the sensor performance.

Due to the collective nature of the interactions the system conserves the total angular momentum, \(\mathbf{\hat S}^2\). It can in this way be conveniently expressed in the Dicke basis \(\left\{ \ket{S, S_z} \right\}\), characterized by the quantum numbers of total angular momentum \(S\) and its projection along a given axis \(S_z\). The allowed values of \(S\) range from \(0\) (or \(1/2\)) up to \(N/2\), depending on the parity of \(N\), while \(S_z\) takes values in the interval \(-S, -S+1, \dots, S\) with the following algebra of collective operators~\cite{Dicke_PhysRev.93.99}:
\begin{equation}
	\label{eq:dicke-algebra}
	\begin{aligned}
		\mathbf{\hat S}^2  \ket{S, S_z} & = S (S + 1) \ket{S, S_z}                                \\
		\hat S_z  \ket{S, S_z}          & = S_z \ket{S, S_z}                                      \\
		\hat S_\pm  \ket{S, S_z}        & = \sqrt{(S \mp S_z) (S \pm S_z + 1)} \ket{S, S_z \pm 1}
	\end{aligned}
\end{equation}

where \(\hat S_{\pm}= \hat S^x \pm i \hat S^y\).

The FTC in this model exhibits period-doubling magnetization dynamics due to an extensive number of low-temperature cat states in its \(\hat H_{\rm LMG}\) spectrum and the induced kicking among them. Specifically, considering the system in its maximum angular momentum sector, \(S = N/2\), once tuned to its ferromagnetic phase \(B < J\) one observes a \(\mathbb{Z}_2\) SSB  over an extensive fraction of the total eigenstates.  In the macroscopic limit \(N \to \infty\), eigenstates with energies below the so-called broken-symmetry edge, \(E^* \equiv -B N\), appear in quasi-degenerate doublets of cat states. Each state in the doublet is localised in the \(\hat{S}_z\) eigenbasis, i.e., \(\ket{S,\pm S_z}\), around either positive or negative magnetization sectors. This localization becomes well-defined as \(N \to \infty\); for finite system sizes, however, eigenstates correspond to even and odd superpositions of the symmetry-broken pair with an exponentially small gap still existing \cite{Dusuel2005}. It is worth remarking that this FTC phenomenology is robust to perturbations, such as an imperfect kicking phase (inducing a rotation \(\phi \neq \pi\), but close to it), other interactions or different initial state preparations.

Within a semiclassical approximation, we compute analytically  the leading order for the overlapping terms (see SM~\cite{SM} Sec.~\ref{sec:sm-semiclassical-lmg}):
\begin{equation}\label{eq:lmg-matrix-elements}
	\frac{| (\hat S_z)_{k\bar{k}}|}{N} \approx \left( \frac{1}{2} - \frac{k}{N} \right) \left(1 - \frac{1}{2}\frac{B^2}{J^2}  - \frac{1}{8}\frac{B^4}{J^4}\right) - \frac{2k+1}{4N}\frac{B^4}{J^4},
\end{equation}
where \(k\) corresponds to the k'th excited pair of cat states in the LMG spectrum. The overlapping terms thus scale extensively with the system size, leading to a quadratic Heisenberg scaling in the QFI. Approaching the phase transition leads to an overall decrease of the QFI, according to \(| (\hat S_z)_{k\bar{k}}| / N \sim (1-(\frac{B}{J})^2)^{1/2}\) for \(B/J<1\) \cite{Botet1982_critical_exp}.

The dynamics for the different initial preparations are shown in Fig.~\ref{fig:qfi-dynamics-lmg}(c), illustrating our discussions.
In a fully polarised initial state, \(|\psi(0)\rangle = |\uparrow ... \uparrow\rangle \) the QFI grows in step-like structures and then decays likewise, on timescales associated with the quasi-degenerate cat state gaps. An initial highly correlated state, \(|\psi(0)\rangle = (|\uparrow ... \uparrow\rangle + |\downarrow ... \downarrow\rangle)/\sqrt{2}\),  on the other hand, shows a high initial QFI at Heisenberg scaling, followed by a sequence of decreasing steps at the same timescales. Lastly, the preparation of the initial state in the ground \(\pi\)-paired subspace leads to a single step-like QFI dynamics, increasing or decreasing depending on the initial correlations in the state, which become proportional to each other for sufficiently long times (inverse of the ground state gap).

\section{Conclusions }
\label{sec:conclusions}

In this work we discussed the performance of general FTCs as quantum sensors of AC fields. Our results show analytically how to exploit their cat state spectral structure, inducing resonant transitions between them in order to maximize the QFI up to the Heisenberg scaling limit.
We have calculated the QFI dynamics for several relevant classes of sensor preparations, both in linear and nonlinear response, unraveling their characteristic step-like dynamics.
We hope our results thus serve as a solid theoretical basis for the potential use of such phases as quantum sensors.

Our theoretical analysis of AC field sensing using FTCs in closed systems is largely platform-independent, allowing its application across a variety of experimental settings. A potential natural realization could be achieved with trapped-ion spin chains~\cite{RevModPhys.93.025001}, where individual ions encode effective spins and their interactions are engineered via optical spin-dependent forces. The capability to prepare correlated many-body states and probe their non-equilibrium dynamics makes such platforms ideally suited for simulating and implementing FTC-based sensors under external AC fields~\cite{Zhang2017}.
Prethermal discrete time crystals also offer a compelling platform. In fact, recent experiments on strongly driven, dipolar-coupled \(C_{13}\) nuclear spins in diamond have demonstrated frequency-selective sensing of AC magnetic fields in the \(0.5-50\) kHz range with no reported system-size-based enhancement in sensitivity~\cite{moonDiscreteTimeCrystal2024}. Together, these developments underscore the ability of implementing FTC-based sensing protocols in near-term setups by leveraging existing quantum control techniques to enhance the precision of AC field measurements.

\section{Acknowledgements}

F.I. acknowledges financial support from the Brazilian funding agencies CAPES, CNPQ, and FAPERJ (No. 151064/2022-9, and No. E-26/201.365/2022), and by the Serrapilheira Institute (grant number Serra – 2211-42166). A.T. acknowledges financial support from the Brazilian funding agency CAPES.

\bibliography{biblio}

\widetext
\clearpage

\begin{center}
	\large \textbf{Supplemental Material} \\ \vspace{0.3cm}
	Andrei Tsypilnikov, Matheus Fibger and Fernando Iemini
\end{center}


In this Supplemental Material (SM), we provide an extended discussion of cat states, further details on the derivation of the QFI and its exact expressions, different AC signals and the initial preparation of the sensor, and more details on the LMG model and its analytical derivations.

\section{SSB and cat states}
\label{sec:sm-ssb-cat}

As mentioned in Section II of the main text, due to SSB a set of the Floquet Hamiltonian eigenvectors (Eq.~\eqref{eq:hf-floquet-schrodinger}) arise as paired states \(\{|E_i\rangle,| E_{\bar{i}}\rangle\}_{i=1}^M\) with opposed parities, where \(M\leq d_H\) and the pair \((i, \bar{i})\) labels the \(i\)-th pair of states. These \(\pi\)-paired eigenstates further take the form of cat states, as we discuss in detail below.

One can first consider the Hermitian operator \(\hat S_z\) anti-commuting with the symmetry operator, as given in Eq.~\eqref{eq:anticomm-x-sz}. Denoting by \(\lambda_i\) and \(|\lambda_i\rangle \) the eigenvalues and eigenstates of the operator \(\hat S_z\), the above relation implies that
\begin{equation}
	\label{eq:sm-parity-condition}
	\hat X |\lambda_i\rangle = |-\lambda_i\rangle.
\end{equation}
that is, the parity operator flips each eigenstate to its counterpart with the opposite eigenvalue. This can be seen from simple algebra:
\begin{eqnarray}
	\hat S_z (\hat X |\lambda_i \rangle ) &=& - \hat X  \hat S_z |\lambda_i \rangle \nonumber \\
	&=& -\lambda_i ( \hat X   |\lambda_i \rangle )
\end{eqnarray}
where, in the first line, we used the anticommutation relation between the operators. From the second line, we can then identify the resulting state as an eigenstate of \(\hat{S}_z\) with a flipped eigenvalue. We can now expand the Floquet eigenstates in this basis,
\begin{equation}
	|E_i \rangle = \sum_{j,\lambda_j>0} c_{ij}^> |\lambda_j\rangle + \sum_{j,\lambda_j<0} c_{ij}^< |\lambda_j\rangle + |\Lambda_i=0\rangle
\end{equation}
where \(|\Lambda_i=0\rangle = \sum_{j,\lambda_j=0} c_{j}^0 |\lambda_j\rangle\) is its decomposition in the null eigenvalue subspace.
Recalling that the Floquet eigenstates are parity eigenstates, in order to fulfill the condition \(\hat X |E_i\rangle = p_i |E_i \rangle\) (i.e. Eq.~\eqref{eq:hf-floquet-schrodinger}), and using Eq.~\eqref{eq:sm-parity-condition}, one must have that
\(c_{ij}^< = p_i c_{ij}^>\)
and \(\hat X |\Lambda_i=0\rangle = p_i |\Lambda_i=0\rangle\). The eigenstates then take the simpler form,
\begin{equation}
	|E_i \rangle = |\Uparrow_i\rangle + p_i |\Downarrow_i\rangle + |\Lambda_i=0\rangle
\end{equation}
with
\begin{eqnarray}
	\label{eq:sm-cat-states-def}
	|\Uparrow_i\rangle &=& \sum_{j,\lambda_j>0} c_{ij}^> |\lambda_j\rangle, \nonumber \\
	|\Downarrow_i\rangle &=& \sum_{j,\lambda_j<0} c_{ij}^< |\lambda_j\rangle,         \\
	|\Lambda_i=0\rangle &=& \sum_{j,\lambda_j=0} c_{j}^0 |\lambda_j\rangle  \nonumber
\end{eqnarray}

The SSB phenomenology assumes that, for sufficiently large system sizes, there are paired states \(|E_i\rangle \) and \(|E_{\bar{i}}\rangle\) which are similar to each other apart from their parities (\(|\Uparrow_i (\Downarrow_i) \rangle  \approx |\Uparrow_{\bar{i}} (\Downarrow_{\bar{i}}) \rangle \)).
Moreover, their dominant terms lie in the non-null eigenvalue subspace (\(|\Lambda_{i(\bar{i})} \rangle \approx 0\)), such that, within a small perturbing field \(\epsilon \hat S_z\) the eigenstates tend to collapse along a preferred ``spontaneously broken'' direction \(|\Uparrow_i\rangle \) or \(|\Downarrow_i\rangle\), thus generating states with a nonzero macroscopic magnetization, \(\langle \Uparrow_i (\Downarrow_i) |\hat S_z | \Uparrow_i (\Downarrow_i) \rangle = O(N)\).

\section{Derivation of the HSO}
\label{sec:sm-derivation-qfi}

We discuss here the derivation of the QFI, by analyzing the structure of its corresponding Heisenberg signal operator. We discuss first the simpler case, of linear response, which allows us to obtain an insightful HSO expansion for general AC fields (Sec.~\ref{subsec:sm-hso-lr}) and results in a leading block-diagonal structure once in PDR (Sec.~\ref{subsec:sm-response-block}). Subsequently we expand the analysis to the nonlinear response case, with the AC field in period doubling resonance along the SSB of the FTC (Sec.~\ref{subsec:sm-nlr-hso}).

\subsection{Linear response}
\label{subsec:sm-hso-lr}

In order to solve the QFI, we first notice that for stroboscopic times in the sensor the Heisenberg signal operator can be decomposed as follows,
\begin{equation}
	\label{eq:sm-qfi-h-sum}
	\hat S_h(nT) = \sum_{k=0}^{n-1} \mathcal{\hat P}_k,
\end{equation}
with each \(\mathcal{\hat P}_k\) representing the contribution from the \(k\)th till \((k+1)\)th kick in the sensor Hamiltonian,
\begin{equation}
	\mathcal{\hat P}_k = \int_{kT}^{(k+1)T} \hat U_h(t')^\dagger \left(
	\frac{\partial \hat H(t')}{ \partial h} \right) \hat U_h(t') dt'
\end{equation}

Defining the operator
\begin{equation}
	\hat u_{h}^{[t_1,t_2]} \equiv \mathcal{T} e^{-i \int_{t_1}^{t_2} (\hat H_F + h f(t') \hat O)dt'},
\end{equation}
since \(\hat{H}_F\) commutes with the parity operator \(\hat X\), the unitary operator can be represented in the form,
\begin{equation}
	\hat U_{h \rightarrow 0}(kT+\Delta t) = \hat u_{0}^{[kT,kT+\Delta t]} \left(  \prod_{m=1}^k \hat {X} u_{0}^{[(m-1)T,mT]} \right)= \hat X^k \,\hat u_{  0}^{[kT,kT+\Delta t]} \left(  \prod_{m=1}^k \hat u_{ 0}^{[(m-1)T,mT]} \right)
\end{equation}

To simplify notation, let us denote,
\begin{equation}
	\mathcal{\hat  U}_{k,h\rightarrow 0} \equiv \left(  \prod_{m=1}^k \hat u_{ 0}^{[(m-1)T,mT]} \right), \quad \mbox{such that} \quad
	\hat U_{h\rightarrow 0}(kT+\Delta t) = \hat X^k \,\hat u_{0}^{[kT,kT+\Delta t]} \, \mathcal{\hat  U}_{k,h\rightarrow 0}
\end{equation}

Using this form in Eq.~\eqref{eq:sm-qfi-h-sum} and the fact that \(\partial \hat H(t) /\partial h = f(t) \hat O\) one obtains,
\begin{equation}
	\hat S_{h \rightarrow 0}(nT) = \sum_{k=0}^{n-1} \, \mathcal{\hat  U}_{k,h\rightarrow 0}^\dagger \left(
	\int_{kT}^{(k+1)T} f(t')
	(\hat u_{0}^{[kT,t']})^\dagger
	(\hat X^\dagger)^k \hat O (\hat X)^k
	\hat u_{0}^{[kT,t']} dt'
	\right) \mathcal{\hat  U}_{k,h\rightarrow 0}
\end{equation}
Expanding the operator in the Floquet eigenbasis \(\hat O = \sum_{ij} O_{ij} |E_i\rangle \langle E_j|\), with \(O_{ij} = \langle E_i|\hat O|E_j\rangle\), and recalling the spectral properties of Eq.~\eqref{eq:hf-floquet-schrodinger} leads to,
\begin{equation}
	\hat S_{h\rightarrow 0}(nT) =  \sum_{i,j} O_{ij}  |E_{i}\rangle \langle E_{j}| \times  \left( \sum_{k=0}^{n-1} (p_i p_j)^k e^{i \Delta_{ij,h} kT}  \left[ \int_{0}^{T} f(kT+t')   e^{i \Delta_{ij} t' }dt'    \right] \right)
\end{equation}
In a simpler form,
\begin{equation}
	\label{eq:sm-sh-nt}
	\hat S_{h\rightarrow 0}(nT) =  \sum_{i,j}O_{ij} R_{ij}(nT)  |E_{i}\rangle \langle E_{j}|,
\end{equation}
where \(R_{ij}(nT)\) corresponds to a weighted geometric progression,
\begin{equation}
	R_{ij}(nT) = \sum_{k=0}^{n-1} g_{k} q_{ij}^k, \quad \mbox{with } \left[ \begin{array}{lll}
		g_{k}  & = & \int_{0}^{T} f(kT+t')   e^{i \Delta_{ij} t' }dt' \\
		q_{ij} & = & (p_i p_j) e^{i \Delta_{ij} T}
	\end{array} \right.
\end{equation}
In this form we see that while the \(O_{{ij}}\) elements, given by the overlap between the eigenstates within the AC field operator, represent the potential of the AC field operator to correlate the different eigenstates of the dynamics, the \(R_{ij}(nT)\) terms relate their time-dependent response to the varying signal.

\subsection{Response terms in PDR and block-diagonal structure}
\label{subsec:sm-response-block}

Once in PDR the FTC sensor can sustain quadratic scaling in time for its HSO norm for exponentially long times in \(N\).
We can see it by first noticing that for time scales \(t \lesssim \Delta_{ij}^{-1}\) one can roughly neglect the gap dephasing contributions (i.e., \(e^{i\Delta_{ij}t} \approx 1)\)), and the corresponding response in PDR can be approximated,
\begin{equation}
	R_{ij}(nT)
	\approx  \int_0^{nT} \left| \frac{(1-p_i p_j)}{2}f(t')-\bar{f} \right| dt', \qquad t \lesssim \Delta_{ij,h}^{-1},
\end{equation}
with  \(\bar{f} = \int_0^T f(t')dt'/T\) the average signal. The above expression maximizes the response,  scaling linearly with time
\( R_{ij}(t) \sim \mathcal{O}(t)\).
For a time longer than this time scale, the gap induces non-trivial dephasing, leading to sublinear growth in time.
In this way, while response terms between non-\(\pi\)-paired states may quickly vanish in time (below their maximum linear growth) due to their finitely small gaps, those corresponding to \(\pi\)-paired states maintain a coherent linear growth up to exponentially long times.
More precisely, one has that
\( R_{i\bar{i}}(t) \sim \mathcal{O}(t)\) for
\(t \lesssim \mathcal{O}(e^{N})\) due to their vanishing gaps, \(\Delta_{i\bar{i}} \sim  \mathcal{O}(e^{-N}) \).
In this way, after an initial transient time \(t \sim \mathcal{O}(\Delta_{i,j\neq (i,\bar{i})}^{-1})\), the HSO is reduced to a block diagonal form along the cat subspaces, as shown in the main text.

\subsection{Nonlinear response: PDR along SSB direction}
\label{subsec:sm-nlr-hso}

We expand here the discussion of the HSO to the case of nonlinear response. In this case  we focus on AC fields described by a Heaviside signal under period doubling resonance with the sensor (\(f(t+T) = - f(t)\) with \(|f(t)| = 1, \forall t\)). Moreover, the AC field is aligned along the underlying SSB of the FTC, i.e. \(\hat O = \hat S_z\). We first notice from the commuting properties of ~\eqref{eq:anticomm-x-sz}, the important property that,
\begin{equation}
	\hat X \hat u_{h}^{[t_1,t_2]} \hat X = \hat u_{-h}^{[t_1,t_2]}
\end{equation}
where the kick operator flips the sign of the modulated signal. In this way we can represent the unitary in the evolution in the form,
\begin{equation}
	\hat U_h(kT+\Delta t) = \,\hat u_{h}^{[kT,kT+\Delta t]} \left(  \prod_{m=1}^k \hat {X} u_h^{[(m-1)T,mT]} \right)= \hat X^k \,\hat u_{(-1)^k h}^{[kT,kT+\Delta t]} \left(  \prod_{m=1}^k \hat u_{(-1)^{(m-1)} h}^{[(m-1)T,mT]} \right)
\end{equation}

To simplify notation, let us denote,
\begin{equation}
	\mathcal{\hat  U}_{k,h} \equiv \left(  \prod_{m=1}^k \hat u_{(-1)^{(m-1)} h}^{[(m-1)T,mT]} \right), \quad \mbox{such that} \quad
	\hat U_h(kT+\Delta t) = \hat X^k \,\hat u_{(-1)^k h}^{[kT,kT+\Delta t]} \, \mathcal{\hat  U}_{k,h}
\end{equation}

Using this form in Eq.~\eqref{eq:sm-qfi-h-sum}, the fact that \(\partial \hat H(t) /\partial h = f(t) \hat S_z\) and the commutation relation of Eq.~\eqref{eq:anticomm-x-sz} one obtains,
\begin{equation}
	\hat S_h(nT) = \sum_{k=0}^{n-1} (-1)^{k}\, \mathcal{\hat  U}_{k,h}^\dagger \left(
	\int_{kT}^{(k+1)T} f(t')
	(\hat u_{(-1)^{k} h}^{[kT,t']})^\dagger
	\hat S_z
	\hat u_{(-1)^{k} h}^{[kT,t']} dt'
	\right) \mathcal{\hat  U}_{k,h}
\end{equation}
where we see that the kicking terms \(\hat X\) give rise to a \((-1)^k\) modulation in the sum.

A further simplification of the above equation arises observing that in all unitary operators \(\hat u_{...}^{[...]}\) in the above decomposition,
the "\((-1)^k\)" modulation in their amplitude induced by the kicking operator \(\hat X\) is canceled due to the AC field modulation itself. In other words, the AC field amplitude is always in phase with the induced kicking flip. In this way the unitary dynamics is driven by an effective Floquet Hamiltonian, \(\hat H_{F,h}^{[\rm effective]} = \hat H_F + h \hat O\), with
\(\hat u_{(-1)^{k} h}^{[kT,t']} = e^{-i\int_{kT}^{t'} (\hat H_F + h \hat S_z )dt' }\), \(t \leq (k+1)T \). Therefore, all unitaries commute among each other,
sharing moreover a set of time-independent eigenstates:
\begin{equation}
	\hat u_{(-1)^k h}^{[kT,t']} |E_{i,h}\rangle = e^{-i E_{i,h} (t'-kT)}|E_{i,h}\rangle,
\end{equation}
with \(t'\leq (k+1)T\), \(\forall k\).
We use the notation \(E_{i,h}\) and \(\Delta_{ij,h} \equiv  E_{i,h}-E_{j,h}\) since these are their natural interpretation on top of the Floquet Hamiltonian spectrum. In the simplest case, of linear response, we recover our previous notation using \(|E_i \rangle \equiv |E_{i,h\rightarrow 0}\rangle\) and
\(\Delta_{ij} \equiv \Delta_{ij,h\rightarrow 0}\).

Expanding the HSO in such an eigenbasis we obtain,
\begin{equation}
	\hat S_h(nT) =  \sum_{i,j} O_{ij,h}  |E_{i,h}\rangle \langle E_{j,h}| \times  \left( \sum_{k=0}^{n-1}(-1)^{k} e^{i \Delta_{ij,h} kT}  \left[ \int_{0}^{T} f(kT+t')   e^{i \Delta_{ij,h} t' }dt'    \right] \right)
\end{equation}
where \(O_{ij,h} = \langle E_{i,h}|\hat O|E_{j,h}\rangle\).
In a simpler form,
\begin{equation}
	\hat S_h(nT) =  \sum_{i,j}O_{ij,h} R_{ij,h}(nT)  |E_{i,h}\rangle \langle E_{j,h}|,
\end{equation}
where \(R_{ij,h}(nT)\) corresponds to a weighted geometric progression,
\begin{equation}
	R_{ij,h}(nT) = \sum_{k=0}^{n-1} g_{k,h} q_{ij,h}^k, \quad \mbox{with } \left[ \begin{array}{lll}
		g_{k,h}  & = & (-1)^k \int_{0}^{T} f(kT+t')   e^{i \Delta_{ij,h} t' }dt' \\
		q_{ij,h} & = & e^{i \Delta_{ij,h} T}
	\end{array} \right.
\end{equation}
As before, we observe that the \(O_{{ij,h}}\) elements represent the potential of the AC field operator to correlate different eigenstates of the dynamics. Meanwhile, the \(R_{ij,h}(nT)\) terms describe how the system responds to a varying signal over time.

Similar to the linear response case, the same arguments regarding the response terms and their leading block-diagonal structure apply here. The only difference is that the timescales for the elements to vanish due to dephasing are now described by the generalized gap, \(\Delta_{ij,h}^{-1}\), which may not be exponentially long in \(N\) depending on the non-null perturbation in the effective Floquet Hamiltonian, \(\hat H_{F,h}^{[\rm effective]}\).

\subsection{Analytical calculation of response terms}

In order to illustrate the above discussions, we compute below the resonance terms analytically in two illustrative cases; namely, a Heaviside and a sinusoidal signal.\\

\textbf{Case 1. Heaviside step signal in period doubling resonance:} Considering explicitly a Heaviside step function in period doubling resonance, \(f(t+kT) = (-1)^k f(t)\) with \(|f(t)| = 1, \forall t\),  we have that

\begin{eqnarray}
	g_{k,h} & = &  (-1)^{k}\int_{0}^{T} (-1)^{k} e^{i \Delta_{ij,h} t' }dt' = i\frac{1-q_{ij,h}}{\Delta_{ij,h}} \qquad \mbox{(independent of k)} \\
	\sum_{k=0}^{n-1} q_{ij,h}^k  &=&  \frac{1-q_{ij,h}^n}{1-q_{ij,h}} \qquad \mbox{(geometric progression sum)}
\end{eqnarray}

Therefore the response term reduces to,
\begin{equation}
	R_{ij,h}(nT) =  i \frac{1-q_{ij}^n}{\Delta_{ij,h}} =
	\frac{ \sin(\Delta_{ij,h}nT/2)}{\Delta_{ij,h}/2} e^{i \Delta_{ij,h}nT/2 }
\end{equation}
where we used that
\((1-e^{i x}) = e^{ix/2}(-2i \sin(x/2))\).\\

\textbf{Case 2. Sinusoidal signal in period doubling resonance:} Considering a general harmonic function \(f(t) = \sin{\left(\frac{\pi t} { T} +\phi_{\mathrm{AC}}\right)}\), we have that

\begin{equation}
	\begin{aligned}
		g_{k,h} & = (-1)^k \int_{0}^{T} {\sin{\left(\frac{\pi t'} {T} + \pi k +\phi_{\mathrm{AC}}\right)} e^{i \Delta_{ij,h} t' }dt'}                                                 \\
		        & = T \frac{\left(1+e^{i \Delta_{ij,h}  T}\right) \left(\pi  \cos{\phi_{\mathrm{AC}}}-i \Delta_{ij,h} T \sin{\phi_{\mathrm{AC}}}\right)}{\pi ^2-\Delta_{ij,h} ^2 T^2}
	\end{aligned}
\end{equation}

The response term then:
\begin{equation}
	\begin{aligned}
		R_{ij,h}(nT) = \kappa(\phi_{\mathrm{AC}}, \Delta_{ij,h}) \frac{  \sin(\Delta_{ij,h}nT/2)}{\Delta_{ij,h}/2} e^{i \Delta_{ij,h}nT/2}
	\end{aligned}
\end{equation}

where
\begin{equation}
	\begin{aligned}
		\kappa(\phi, \Delta) & \equiv \frac{\pi \Delta T }{\tan{\left(\Delta T / 2\right)} \left( \pi ^2-\Delta ^2 T^2 \right)} \left[ \left(\frac{1}{2}+\frac{T\Delta}{2\pi}\right) e^{i \phi}+\left(\frac{1}{2}-\frac{T\Delta}{2\pi}\right)e^{-i \phi} \right] \\
		                     & = \frac{2 \cos{\phi}}{\pi}+i \frac{2 \Delta  T \sin{\phi}}{\pi^2}+O\left(\Delta ^2\right)
	\end{aligned}
\end{equation}

\section{Characteristic step-like QFI dynamics}
\label{sec:sm-step-like-dynamics}
Under linear response and PDR, each of the block terms of the Heisenberg signal operator (Eq.~\eqref{eq:hso-block-diagonal}) grows linearly in time for times smaller than its corresponding gap, \(\hat s^{i \bar{i}}(t \ll \Delta_{i \bar{i}}^{-1}) \sim t \hat s^{i \bar{i}}(0)\), while having a sublinear dynamics for longer times and eventually vanishing on top of the standard quantum limit, \(\hat s^{i \bar{i}}(t \gg \Delta_{i \bar{i}}^{-1})/t \sim \hat 0\). Therefore, the QFI renormalized by the standard quantum limit scaling, \(F_h(t)/Nt^2\), will generally exhibit a step-like behavior on these time scales (\(t \sim \Delta_{i \bar{i}}^{-1}\)) due to the successive vanishing of the block diagonal terms. Specifically, from Eq.~\eqref{eq:hso-block-diagonal} the QFI is given by,
\begin{equation}
	\label{eq:sm-qfi-block-diag}
	F_h(t) = \sum_i \langle (\hat s^{i \bar{i}})^2 \rangle - (\sum_i \langle \hat s^{i \bar{i}} \rangle)^2
\end{equation}
and the step difference of Eq.~\eqref{eq:step-qfi-definition} reduces to,
\begin{equation}
	\label{eq:sm-step-qfi-bd}
	\delta_k F_h = -\langle \hat s^{i \bar{i}} (0)^2 \rangle + \langle \hat s^{i \bar{i}} (0) \rangle (2 \langle  \hat S_{h,\rm bd}(t \ll \Delta_{k \bar{k}}^{-1} ) \rangle - \langle \hat s^{i \bar{i}} (0) \rangle )
\end{equation}
The steps are therefore  determined by the initial preparation of the sensor and the magnetization along the AC field direction before the corresponding characteristic time.
Defining the projector in the cat subspace, \(\hat P_{\rm cat} = \sum_{i=1}^M |E_i\rangle \langle E_i|\), while the orthogonal subspace to it has a vanishing contribution to the HSO, the projection of the initial state on such subspace \(\hat P_{\rm cat}|\psi(0)\rangle\), with an overlap \(p_{\rm cat} = \langle \psi(0)| \hat P_{\rm cat} |\psi(0)\rangle\),  dictates the long-time QFI dynamics.\\

\subsection{Symmetry breaking initial states}
\label{sec:sm-symmetry-breaking}
We consider here the symmetry breaking initial states as defined in Eq.~\eqref{eq:init-ssb}.
For sufficiently large system sizes, the broken symmetry states \(|\Uparrow_i (\Downarrow_i) \rangle \sim
	|\Uparrow_{\bar{i}} (\Downarrow_{\bar{i}}) \rangle\) where the corresponding pairs of cat states differ only in their parity. Consequently, one has that
\(|\Uparrow_i \rangle \sim (|E_i \rangle + |E_{\bar{i}}\rangle)/\sqrt{2}\). In this way \(\langle \hat s^{i\bar{i}} (0)^2 \rangle = c_i |O_{i\bar{i}}|^2\) and
\(\langle \hat s^{i\bar{i}} (0) \rangle = c_i \Re(O_{i\bar{i}})\). The
step difference of Eq.~\eqref{eq:sm-step-qfi-bd} is given by,

\begin{equation}
	\delta_k F_h = c_i\Re(O_{i\bar{i}}) \left(2 \sum_{j \in M_k} c_j \Re(O_{j\bar{j}}) - c_i \Re(O_{i\bar{i}})\right) - c_i |O_{i\bar{i}}|^2
\end{equation}

where we define \(M_k\) as the set of cat-state pairs with a gap smaller (or equal) than the k'th pair, i.e., which have a larger lifetime in the HSO as compared to the k'th pair subspace. Precisely, \(M_k = (i_1,i_2,...,i_{k'})\) with indexes such that \(\Delta_{i_j,\bar{i}_j} \leq  \Delta_{i_k,\bar{i}_k}\). In order to obtain a clearer picture of the steps, we make a reasonable approximation as follows. Since the overlapping terms in the above equation all involve cat states, and therefore scale with the number of spins, we assume they are qualitatively the same for sufficiently large systems, \(O_{i\bar{i}} = O_{j\bar{j}} \equiv  o \) and assume them to be real numbers (\(o \in \mathcal{R} \)) for simplicity. The step difference in this way reduces to,
\begin{equation}
	\delta_k F_h = o^2 c_k (\mathcal{P}_{M_k} - \mathcal{P}_{\perp M_k})
\end{equation}
where \(\mathcal{P}_{M_k} = \sum_{i \in M_k} c_i\) is the overlap of the initial state within the \(M_k\) subspace, and
\(\mathcal{P}_{\perp M_k} = \sum_{i \in M, i \notin M_k} c_i\) its overlap in the cat subspace, but out of \(M_k\). The QFI dynamics is thus characterized by a sequence of increasing steps - while \(\mathcal{P}_{M_{[...]}} > \mathcal{P}_{\perp M_{[...]}}\) - and eventually a further sequence of decreasing steps, as discussed in the main manuscript.

\subsection{Initial states in a single cat subspace}
\label{sec:sm-single-subspace}
In the case of an initial state represented in the form,
\begin{equation}
	|\psi_{i\bar{i}}(0)\rangle = c_i |E_i\rangle + c_{\bar{i}} |E_{\bar{i}}\rangle
\end{equation}
substituting such an initial state in Eq.~\eqref{eq:sm-sh-nt} one obtains
\begin{equation}
	\begin{aligned}
		\langle \psi_{i\bar{i}}(0) |\hat S_{h \rightarrow 0}(nT) |\psi_{i\bar{i}} (0)\rangle                                    & = \hat O_{i\bar{i}}
		\bigg(c_i^* \langle E_i| + c^*_{\bar{i}} \langle E_{\bar{i}}|\bigg)
		\bigg( R_{i\bar{i}}(nT) |E_i\rangle \langle E_{\bar{i}}| + R^*_{i\bar{i}}(nT)|E_{\bar{i}}\rangle \ \langle E_i| \bigg)
		\bigg(c_i |E_i \rangle + c_{\bar{i}} |E_{\bar{i}} \rangle \bigg)                                                                                                                                                                                                                                                             \\
		                                                                                                                        & = \left(c^*_i  c_{\bar{i}}  O_{i\bar{i}} R_{i\bar{i}}(nT)  +  c_i c^*_{\bar{i}} O^*_{i\bar{i}}   R^*_{i\bar{i}}(nT) \right) = 2\Re{ \left(c^*_i c_{\bar{i}}  O_{i\bar{i}} R_{i\bar{i}}(nT)\right)} \\
		\langle \psi_{i\bar{i}}(0) |\hat S^\dagger_{h\rightarrow 0}(nT) \hat S_{h\rightarrow 0}(nT) |\psi_{i\bar{i}} (0)\rangle & = \left(|c_i|^2 + |c_{\bar{i}}|^2 \right) | O_{i\bar{i}} R_{i\bar{i}}(nT)|^2
	\end{aligned}
\end{equation}

The QFI reduces to the form,
\begin{equation*}
	F_h(nT) = 4 \left[ \left( |c_i|^2 + |c_{\bar{i}}|^2 \right) \left| O_{i\bar{i}} R_{i\bar{i}}(nT) \right|^2
		- 4 \left| \Re \left( c_i^* c_{\bar{i}} \, O_{i\bar{i}} \, R_{i\bar{i}}(nT) \right) \right|^2 \right],
\end{equation*}
which, in the particular case where \(c_i = \cos(\theta/2)\) and \(c_{\bar{i}} = e^{i\varphi} \sin(\theta/2)\), and using the fact that \(O_{i\bar{i}} \in \mathbb{R}\), yields the compact expression

\begin{equation*}
	\frac{F_h(nT)}{|nT O_{i\bar{i}}|^2} = 4 \left[ \left| \kappa(\phi_{\mathrm{AC}}, \Delta_{i\bar{i}}) e^{i \Delta_{i\bar{i}}nT/2} \right|^2 - \sin^2\theta \left( \Re \left[ e^{i\left(\varphi + \Delta_{i\bar{i}}nT/2\right)} \kappa(\phi_{\mathrm{AC}}, \Delta_{i\bar{i}}) \right] \right)^2 \right] \left(\frac{ \sin(\Delta_{i\bar{i}}nT/2)}{\Delta_{i\bar{i}}nT/2}\right)^2
\end{equation*}

or alternatively:

\begin{equation}
	\label{eq:sm-qfi-general}
	\frac{F_h(nT)}{|nT O_{i\bar{i}}|^2} = C_h \left(1
	- \sin^2 \theta \cos^2{\left(\varphi + \Delta_{i\bar{i}} nT/2 \right)}\right) \left(\frac{ \sin(\Delta_{i\bar{i}}nT/2)}{\Delta_{i\bar{i}}nT/2}\right)^2
\end{equation}

where \(C_h = 4\) for a Heaviside step signal and \(C_h \approx \frac{16 \cos^2{\phi_{\mathrm{AC}}}}{\pi^2}\) for a sinusoidal signal.

\textit{Maximising the time-dependent sensitivity of FTC sensors for symmetry-breaking initial state.}~The term of Eq.~\eqref{eq:sm-qfi-general} encapsulating the temporal evolution is given by
\begin{equation}\label{eq:sm-fq-func}
	f_h(\tau) = \Big(1 - \sin^2{\theta} \cos^2 \left(\varphi+\tau/2  \right) \Big) \frac{\sin ^2\left(\tau / 2\right)}{(\tau / 2)^2}
\end{equation}
with \(\tau = \Delta_{i\bar{i}}t\).
Our goal is to identify the conditions maximizing \(f_h\), thereby optimizing the sensor's precision, and we can see that maximum  achieved for a maximally entangled initial state (\(\theta = 0, \varphi = 0\)) at point \(\tau=0\) and decay at time close to \(\pi\).

For initial states that break the symmetry (\(\theta = \pi/2\)), the occurrence of a maximum within \(0 \le \tau \le 2\pi\) is determined by the relative phase. To determine the maximum of the QFI in such a case, we compute the first derivative with respect to \(\tau\) and set it to zero. After a few simplifications, this yields the transcendental equation:
\begin{equation}
	\label{eq:sm-solve-fq}
	\tau \sin(\tau + \varphi) + \cos(\tau + \varphi) = \cos\varphi.
\end{equation}

\begin{figure}[ht]
	\includegraphics[width=0.6\textwidth]{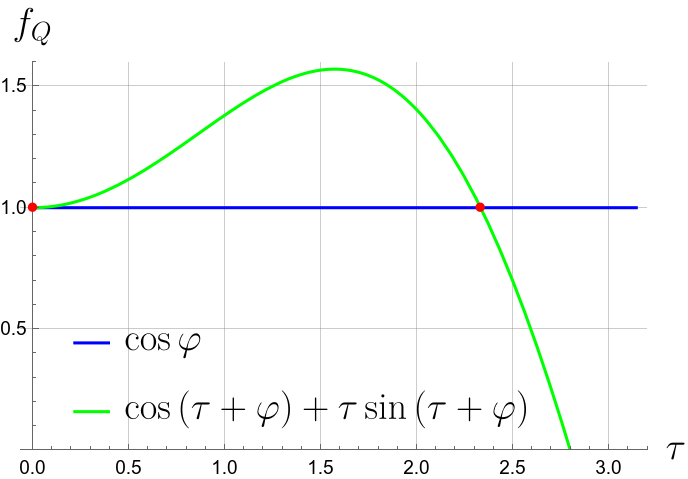}
	\caption{Numerical solution of the transcendental equation (Eq.~\eqref{eq:sm-solve-fq}), derived from the derivative of \(f_h(t)\) (Eq.~\eqref{eq:sm-fq-func}) for the physical state with \(\varphi = 0\). The first non-trivial root, \(\tau_{\text{max}} \approx 2.3311\) where the QFI reaches its maximum value for symmetry-breaking initial state. For \(0 < \varphi < \pi/2\), the root \(\tau_{\text{max}}\) shifts slightly to smaller value and to bigger value of \(\max{f_h}\), indicating a dependence of the QFI peak on the relative phase of the initial state superposition.}
	\label{fig:sm-solution-graph}
\end{figure}

This equation depends on the relative phase \( \varphi \). In the case of \( \varphi = 0 \), corresponding to SSB state \( |\psi_0\rangle = (|E_i\rangle + |E_{\bar{i}}\rangle)/\sqrt{2} \), the value of QFI for this max point could be obtained numerically (see Fig.~\ref{fig:sm-solution-graph}). The first non-trivial root occurs at \( \tau_{\text{max}} \approx 2.331 \), with \( f_h(\tau_{\text{max}}) \approx 0.525 \). Thus, for a SSB state with \( \theta = \pi/2 \) and \( \varphi = 0 \), the peak of the QFI is
\begin{equation}
	\frac{F_h(t_{\text{max}})}{(t_{\text{max}} |O_{i\bar{i}}|)^2} \approx 0.525 \cdot C_h,
\end{equation}
achieved at the optimal sensing time \( t_{\text{max}} = \tau_{\text{max}} / \Delta_{i\bar{i}} \approx  2.331 / \Delta_{i\bar{i}} \).
This value, roughly half of the maximally correlated state, underscores the FTC sensor's capability to leverage Floquet-driven coherence, even from initially low-correlation states, offering a robust platform for quantum sensing applications.

\section{Diagonalization of the effective Floquet Hamiltonian }
\label{sec:sm-effective-floquet}
The HSO within the nonlinear response can be expanded in terms of the eigenstates and eigenvalues of an effective Floquet Hamiltonian, \(\hat H_{F,h}^{\rm[effective]} = \hat H_F + h \hat O\), which is the bare Floquet Hamiltonian with a constant perturbation from the AC field operator. We are interested in the case where the perturbation explicitly breaks the Hamiltonian symmetry, i.e., \(\hat O=\hat S_z\). Expanding the effective Hamiltonian in the unperturbed Hamiltonian basis of a single cat subspace, with eigenvalues \(\{E_{i,h=0},E_{\bar{i},h=0}\}\) and eigenvectors \(\{|E_{i,h=0}\rangle, |E_{\bar{i},h=0}\rangle\}\), we have that
\begin{equation}
	\left( \hat H_F + h \hat S_z\right)_{i\bar{i}} = \left(
	\begin{array}{cc}
			E_{i,0} & h z_{i\bar{i},0} \\ (hz_{i\bar{i},0})^* & E_{\bar{i},0}
		\end{array} \right)
\end{equation}
where \(z_{i\bar{i},h}=\langle E_{i,h} |\hat S_z |E_{\bar{i},h} \rangle\). Given that \(h z_{i\bar{i},0} \neq 0\), the eigendecomposition is computed with corresponding eigenvalues,
\begin{equation}
	E_{i,h} = \frac{(E_{i,0} + E_{\bar{i},0}) + \sqrt{\Delta_{i\bar{i},0}^2 + 4|hz_{i\bar{i},0}|^2}}{2}, \qquad E_{\bar{i},h} = \frac{(E_{i,0} + E_{\bar{i},0}) - \sqrt{\Delta_{i\bar{i},0}^2 + 4|hz_{i\bar{i},0}|^2}}{2},  \qquad
	\Delta_{i\bar{i},h} = E_{i,h} - E_{\bar{i},h}
\end{equation}
and eigenvectors,
\begin{equation}
	\label{eq:sm-perturbed-eigenstates}
	|E_{i,h}\rangle = \left(
	\begin{array}{cc}
			\cos(\theta) \\ e^{i \chi} \sin(\theta)
		\end{array}
	\right), \qquad
	|E_{\bar{i},h}\rangle = \left(
	\begin{array}{cc}
			-e^{-i \chi} \sin(\theta) \\  \cos(\theta)
		\end{array}
	\right)
\end{equation}
where  \(0\leq \theta < \pi/2\), \(0 \leq \chi < 2\pi\),
\begin{equation}
	\label{eq:sm-theta-chi}
	\theta = \frac{1}{2} \atan(\frac{|hz_{i\bar{i},0}|}{\Delta_{i\bar{i},0}}),\qquad \chi = 2\pi - \nu,
\end{equation}
with \(hz_{i\bar{i},0}=|hz_{i\bar{i},0}|e^{i\nu}\).
Recalling that \(\hat S_z = z_{i\bar{i},0} |E_{i,0}\rangle \langle E_{\bar{i},0}| + z_{i\bar{i},0}^* |E_{\bar{i},0}\rangle \langle E_{i,0}|\), along the new basis \(\{|E_{i,h}\rangle,|E_{\bar{i},h}\rangle\}\) its elements are given by,
\begin{eqnarray}
	(\hat S_z)_{mn,h}  \equiv z_{mn,h} =z_{i\bar{i},0} \langle E_{m,h} | E_{i,0} \rangle \langle E_{\bar{i},0} | E_{n,h} \rangle +
	z_{i\bar{i},0}^* \langle E_{m,h} | E_{\bar{i},0} \rangle \langle E_{i,0} | E_{n,h} \rangle
\end{eqnarray}
Therefore, using Eq.~\eqref{eq:sm-perturbed-eigenstates} and the above equation one obtains that in the perturbed basis,
\begin{equation}
	\hat S_z = \left( \begin{array}{cc}
		z_{ii,h} & z_{i\bar{i},h} \\ z_{i\bar{i},h}^* & z_{\bar{i}\bar{i},h}
	\end{array} \right) =
	\left( \begin{array}{cc}
		\Re(z_{i\bar{i},0}  \sin(2\theta) e^{i\chi}) & z_{i\bar{i},0} \cos^2(\theta) - e^{-2i\chi}\sin^2(\theta) \\  H.c & -\Re(z_{i\bar{i},0}  \sin(2\theta) e^{i\chi})
	\end{array} \right)
\end{equation}

A few limiting cases are worth mentioning. If the perturbation is too small as compared to the gap, \(hz_{i\bar{i},0} \ll \Delta_{i\bar{i},0}\), the eigenstates are barely changed with \(\theta \sim 0\) and \(|E_{i,h}\rangle \sim |E_{i,0}\rangle\). At the other extreme, if the perturbations are large as compared to the gap, \(hz_{i\bar{i},0} \gg \Delta_{i\bar{i},0}\), the eigenstates have maximum superposition with \(\theta \sim \pi/4\) and \(|E_{i,h}\rangle \sim |E_{i,0}\rangle \pm |E_{\bar{i},0}\rangle\). Recalling that the unperturbed eigenstates are highly correlated cat states, this superposition then favors their low-correlated parts.

\section{Nonlinear response}
\label{sec:sm-nonlinear-response}
We consider here larger AC amplitude fields, given by a Heaviside signal in PDR to the sensor, \(f(t+T) = -f(t)\) with \(|f(t)|=1\), which facilitates an analytical treatment for the QFI (since it allows a time-independent eigendecomposition for the stroboscopic unitary operators in the HSO expansion). The results, however, shall be qualitatively similar for other (non-Heaviside) signals as long as in PDR, following the same reasoning as in the linear response case.

The HSO has the same structure as Eq.~\eqref{eq:qfi-linear-response} with the modified eigendecomposition of Eq.~\eqref{eq:nlr-shift-notation}.
Some important effects are worth mentioning:
(i) while the diagonal terms in \(\hat S_h(t)\) were null in the linear response, this may no longer be the case since the eigenstates explicitly break the symmetry, leading to non-null overlaps \( O_{ii,h}  \neq 0\); (ii) the perturbation opens the \(\pi\)-pairing gaps, so the dephasing it induces in the resonance terms will always occur in finite time \(t\sim \Delta_{i\bar{i},h}^{-1} \leq \Delta_{i\bar{i},h=0}^{-1}\), and will not diverge in the macroscopic limit.

In order to study analytically the NLR, we consider the case of a  field that weakly perturbs the full spectrum, although with significant consequences in the cat subspaces. Specifically, we consider the case of a field satisfying Eq.~\eqref{eq:nlr-shift-notation}.
The analysis is then based on a perturbation theory within the cat subspaces.

One first notices that,
similar to Eq.~\eqref{eq:hso-block-diagonal}, after an initial transient time  \(t \sim \Delta_{ij\neq(i ,\bar{i}),h}^{-1}\) the HSO recovers a block diagonal form,

\begin{equation}
	\label{eq:sm-hso-block-diagonal-nlr}
	\hat{S}_h(t) \approx \hat{S}_{h,\mathrm{bd}}(t) =
	\begin{pmatrix}
		\hat{s}^{[i_1 \bar{i}_1,h]}(t) & 0      & \hdots                         & 0         \\
		0                              & \ddots &                                & \vdots    \\
		\vdots                         &        & \hat{s}^{[i_M \bar{i}_M,h]}(t) & 0         \\
		0                              & \cdots & 0                              & t \hat{D}
	\end{pmatrix}.
\end{equation}
but with nonzero diagonal along the entire spectrum \(\hat D = \mbox{diag}(O_{(M+1)(M+1),h},...,O_{dd,h})\),  and

\begin{equation}
	s^{ [i \bar{i},h] }(t) =
	\left( \begin{array}{cc}
			t O_{ii,h}                       & R_{i\bar{i},h}(t) O_{i\bar{i},h} \\
			R_{\bar{i}i,h}(t) O_{\bar{i}i,h} & t O_{\bar{i}\bar{i},h}
		\end{array} \right)
\end{equation}

Recalling that
the resonant term \(R_{i\bar{i},h}(t)\) is an oscillating sine function with amplitude \(\Delta_{i\bar{i},h}^{-1}\), the diagonal terms become prominent only when they exceed such an amplitude. I.e. as soon as
\(t O_{i i,h} > \Delta_{i\bar{i},h}^{-1} O_{i \bar{i},h}\), which occurs at \(t_{\rm{NLR},i} \sim \Delta_{i\bar{i},h}^{-1} O_{i \bar{i},h}/O_{i i,h} \).
While before this characteristic time scale the HO is a block off-diagonal matrix, with a dynamical behavior qualitatively similar to the LR,
after this time the HO resembles a diagonal Pauli operator, \(s^{ [i \bar{i},h] }(t) \approx  t O_{i\bar{i},o}q_h \hat \sigma_z\) with a constant and unbounded linear scaling in time. Thus, while for \(t \lesssim t_{\rm{NLR},i}\) the corresponding QFI resembles the LR case, for larger times \(t \gtrsim t_{\rm{NLR},i}\) it features a constant quadratic scaling in time.

The overlapping terms can be further determined from the diagonalization of the effective Floquet Hamiltonian in each separate cat subspace, due to the energy scale separation of Eq.~\eqref{eq:pt-condition}. We obtain that (see Sec.~\ref{sec:sm-effective-floquet}),
\begin{equation}
	t_{\rm{NLR},i} = \Delta_{i\bar{i},h}^{-1} \frac{\cos^2(\theta) - e^{-2i\chi}\sin^2(\theta)}{2\sin(2\theta)\cos(\chi)}
\end{equation}
with  \(\theta \) and \(\chi\) given by Eq.~\eqref{eq:sm-theta-chi}. Assuming for simplicity that \(\chi = 0\), the above relation can be rewritten as (using that \(\cos^2(\theta)-\sin^2(\theta) = \cos(2\theta)\)),
\begin{eqnarray}
	t_{\rm{NLR},i} = \frac{1}{2h |O_{i\bar{i},0}|} (\Delta_{i\bar{i},0 } / \Delta_{i\bar{i},h })
	= \left( 2h |O_{i\bar{i},0}| \sqrt{1+ \left(\frac{2h |O_{i\bar{i},0}|}{\Delta_{i\bar{i},0} }\right)^2 } \right)^{-1}
\end{eqnarray}
We see that, apart from corrections of the order of \(2h |O_{i\bar{i},0}|/\Delta_{i\bar{i},0}\), the NLR with its characteristic quadratic growth in the QFI settles into the dynamics at times \(t\sim 1/2h |O_{i\bar{i},0}|\). Two interesting cases are worth computing explicitly, those of a weak or strong perturbation as compared to the unperturbed gap. In the case of a weak perturbation,  \(h |O_{i\bar{i},0}|\ll  \Delta_{i\bar{i},0}\)  the spectral properties of the corresponding subspace are roughly unaffected, and for short times the dynamics shall be similar to the LR case. In fact, \( \Delta_{i\bar{i},h} \approx  \Delta_{i\bar{i},0}\) and \(t_{\rm NLR,i} \approx 1/2h |O_{i\bar{i},0}| \gg 1/ \Delta_{i\bar{i},0} = t_{\rm LR,i}\). The QFI behaves as LR for times much longer than the dephasing time of the subspace. On the other hand, for strong perturbations
\(h |O_{i\bar{i},0}|\gg  \Delta_{i\bar{i},0}\), we have that
\( \Delta_{i\bar{i},h} \approx 2h |O_{i\bar{i},0}|\) and in this way
\(t_{\rm NLR,i} \approx \Delta_{i\bar{i},0}/4h^2 |O_{i\bar{i},0}|^2 \ll 1/2h |O_{i\bar{i},0}| \ll 1/\Delta_{i\bar{i},0}  = t_{\rm LR,i}\). For times much shorter than the dephasing time of the subspace, the NLR sets in the QFI, leading to its characteristic quadratic growth.

\section{Semiclassical approach to LMG model}
\label{sec:sm-semiclassical-lmg}
In order to compute the overlapping terms in the model, we  employ a Holstein-Primakoff expansion around the semiclassical symmetry broken states \cite{Dusuel2005}. The approach allows us to obtain the to obtain analytically the first order corrections of the overlapping terms.

\subsection{Classical energy}
The classical energy should be minimal for the system and it could be obtained by substituting in the bare Hamiltonian \(\hat H_{\text{LMG}}\) the expression for average spin,
\begin{equation}
	\label{eq:sm-rotation-sclass}
	\langle \hat S \rangle = \frac{N}{2}
	\begin{pmatrix}
		\sin{\vartheta} \cos\varphi \\
		\sin{\vartheta} \sin\varphi \\
		\cos{\vartheta}             \\
	\end{pmatrix},
	\begin{array}
		c 0 \le \varphi < 2 \pi
		\\ 0 \le \vartheta \le \pi
	\end{array}
\end{equation}
which gives the  classical energy per spin \(e_c\) as:
\begin{equation}
	\label{eq:sm-classical-energy}
	e_c = \frac{\langle \hat H_0 \rangle}{N} = - \frac{J \cos^2{\vartheta}}{2} - B \sin{\vartheta} \cos{\varphi}
\end{equation}

From the corresponding Euler–Lagrange equation, one finds that the energy is minimized for \(\varphi = 0\), eliminating the second variable in the energy functional. Minimization with respect to \(\vartheta\) yields two distinct regimes:
\begin{equation}
	\label{eq:sm-magnetization-level}
	\begin{aligned}
		\vartheta & = \arcsin{\frac{B}{J}} \quad \text{or} \quad \vartheta = \pi - \arcsin{\frac{B}{J}}, & \quad 0 \le B < J, \\
		\vartheta & = \frac{\pi}{2},                                                                     & \quad B \ge J.
	\end{aligned}
\end{equation}
For \(B \ge J\), the ground state is unique and fully polarized along the direction of the magnetic field. In contrast, for \(0 \le B < J\), the ground state has a twofold degeneracy.

\subsubsection{Semi-classical approach}
The main idea of the semi-classical approach is to consider an expansion in \(\frac{1}{N}\) around the order parameter. The transition to this local basis could be given by the rotation matrix, generally written as \(\hat U_R = e^{-\gamma_1 L_3}e^{-\gamma_2 L_2}e^{-\gamma_3 L_3}\) with generators of the \(SO(3)\) group \(L_1,L_2,L_3\) and Euler angles \(\gamma_1, \gamma_2,\gamma_3\).

The generators of the \( SO(3) \) Lie algebra are given by the following real, antisymmetric \(3 \times 3\) matrices:
\begin{equation}
	L_1 =
	\begin{pmatrix}
		0 & 0 & 0  \\
		0 & 0 & -1 \\
		0 & 1 & 0
	\end{pmatrix}, \quad
	L_2 =
	\begin{pmatrix}
		0  & 0 & 1 \\
		0  & 0 & 0 \\
		-1 & 0 & 0
	\end{pmatrix}, \quad
	L_3 =
	\begin{pmatrix}
		0 & -1 & 0 \\
		1 & 0  & 0 \\
		0 & 0  & 0
	\end{pmatrix}
\end{equation}

These satisfy the commutation relations:
\[
	[L_i, L_j] = \varepsilon_{ijk} L_k
\]
\begin{equation}
	\label{eq:sm-sprime-rotation}
	\hat{S'} =  \hat U_R \hat {S}
\end{equation}

The choice of variables as \(\gamma_1 = \frac{\pi}{2} - \varphi, \quad \gamma_2=\vartheta, \gamma_3 = - \frac{\pi}{2}\) corresponds to the rotation given in Eq. \eqref{eq:sm-rotation-sclass}, resulting inverse matrix (i.e. \(\hat{S} = \hat{U}_R^{-1} \hat {S}' \)) is:
\begin{equation}
	\hat{U}_R^{-1} =
	\left(
	\begin{array}{ccc}
			\cos \vartheta \cos \varphi & -\sin\varphi & \sin\vartheta \cos\varphi \\
			\cos \vartheta \sin \varphi & \cos\varphi  & \sin\vartheta \sin\varphi \\
			-\sin\vartheta              & 0            & \cos\vartheta
		\end{array}
	\right).
\end{equation}
The Euler angle \(\gamma\) is arbitrary and is not defined from the equation on minimum of the classical energy Eq. \eqref{eq:sm-classical-energy}.
The Holstein-Primakoff representation \cite{holstein1940} for quantum spins used to get corrections to the Hamiltonian is expressed by the boson creation and annihilation operators. In that representation, the spin-\(\hat S\) preserves the spin commutation relations (\([\hat{S}_\alpha',\hat{S}_\beta'] = i \epsilon_{\alpha \beta \gamma}\hat{S}_\gamma'\)) for maximal spin \(S=\frac{N}{2}\) it is written as:
\begin{equation}
	\label{eq:sm-holstein-primakoff}
	\begin{aligned}
		\hat{S}_{z}' & =\frac{N}{2}-a^\dagger a                                                                                                \\
		\hat{S}_{+}' & = \sqrt{N} \left( 1 - \frac{1}{2N} a^\dagger a + O(N^{-2})\right)a \approx \sqrt{N}a + O(N^{-1/2})                      \\
		\hat{S}_{-}' & = \sqrt{N} a^{\dagger}\left(1 - \frac{1}{2N} a^{\dagger} a + O(N^{-2}) \right) \approx \sqrt{N}a^\dagger  + O(N^{-1/2})
	\end{aligned}
\end{equation}

Substituting the value for spin Eq.~\eqref{eq:sm-sprime-rotation} in terms of \(\hat S'\) into \(\hat H_{\text{LMG}}\) and using the representation of Eq.~\eqref{eq:sm-holstein-primakoff} one obtains:
\begin{equation}
	\label{eq:HamiltonianBare}
	\hat H_\text{LMG} = N e_c + \delta_b + \epsilon_b  \left(\hat a^{\dagger} \hat a + \hat a \hat a^{\dagger}\right) + \Delta_b \left(\hat a^{\dagger} \hat
	a^{\dagger} + \hat a \hat a \right) + O(N^{-1/2})
\end{equation}
where the coefficients of the bare Hamiltonian are,
\begin{equation}\label{eq:sm-hamiltonian-bare}
	\begin{aligned}
		e_c        & = - \frac{B^2}{2 J} - \frac{J}{2} \\
		\delta_b   & = -J                              \\
		\epsilon_b & = J - \frac{B^2}{2 J}             \\
		\Delta_b   & = -\frac{B^2}{2 J}
	\end{aligned}
\end{equation}

It should be noted here that the Hamiltonian contains no terms proportional to \(\hat  a^\dagger\) or \(\hat a\), i.e., of order \(N^{1/2}\). These terms simply cancel because the angle of the rotation Eq. \eqref{eq:sm-magnetization-level} has been chosen to bring \(\hat{S}_z'\) along the classical magnetization.

\subsubsection{The Bogoliubov transformation}

Equation \eqref{eq:HamiltonianBare} could be rewritten using diagonalized by the Bogoliubov transformation for bosonic operators \(\hat a\) and \(\hat a^\dagger\) is given by:
\begin{eqnarray}
	\hat{a} &=& u \hat{b} + v \hat{b}^\dagger, \nonumber \\
	\hat{a}^\dagger &=& u^* \hat{b}^\dagger + v^* \hat{b},
\end{eqnarray}
where \(u\) and \(v\) are complex coefficients satisfying \(|u|^2 - |v|^2 = 1\) to preserve the bosonic commutation relations.
It can be shown~\cite{Dusuel2005} that if we choose \(u=\cosh{\frac{\Theta}{2}}\) or \(v=\sinh\frac{\Theta}{2}\) with \(\Theta\) corresponding to \(\tanh{\Theta} = - \frac{\Delta_b}{\epsilon_b} = -\frac{B^2}{B^2 + J^2}\). The operators \(\hat S'\) in terms of new operators are:
\begin{equation}
	\begin{aligned}
		\hat{S}_x' & = \frac{1}{2}\left( \hat{S}_+'+ \hat{S}_-'\right) = \frac{\sqrt{N} e^{\frac{\Theta}{2}}}{2} \left(b + b^\dagger \right) + O(N^{-1/2})
		\\
		\hat{S}_y' & = \frac{1}{2i}\left( \hat{S}_+'-\hat{S}_-'\right) = \frac{\sqrt{N} e^{-\frac{\Theta}{2}}}{2i} \left(b - b^\dagger \right) + O(N^{-1/2})                                                    \\
		\hat{S}_z' & = \frac{N}{2} - \cosh^2{\frac{\Theta}{2}} b^\dagger b - \sinh^2{\frac{\Theta}{2}} b  b^\dagger - \sinh {\frac{\Theta}{2}} \cosh {\frac{\Theta}{2}} \left(b^\dagger b^\dagger + b b \right) \\
		           & = \frac{N}{2} + \frac{1}{2} \left(1 - \cosh{\Theta} \right) -  b^\dagger b \cosh{\Theta} - \frac{\sinh{\Theta}}{2}  \left(b^\dagger b^\dagger + b b \right)
	\end{aligned}
\end{equation}
Using the following formulas to represent some expressions
of \(\tanh{\Theta}\):
\begin{equation}
	\begin{aligned}
		e^{\frac{\Theta}{2}} & = \left(\frac{1+\tanh{\Theta}}{1-\tanh{\Theta}}\right)^{\frac 1 4}                                          \\
		\cosh{\Theta}        & = \sqrt{\frac{1}{1-\tanh^2{\Theta}}} ,\quad  \sinh{\Theta} = \frac{\tanh{\Theta}}{\sqrt{1-\tanh^2{\Theta}}}
	\end{aligned}
\end{equation}
we obtain that,
\begin{equation}
	\begin{aligned}
		\hat{S}_x' & = \frac{\sqrt{N}}{2} \left(1 + 2 (B/J)^2 \right)^{-1/4} \left( \hat b + \hat b^\dagger \right) + O\left(N^{-3/2}\right)
		\\
		\hat{S}_y' & = \frac{\sqrt{N}}{2 i} \left(1 + 2 (B/J)^2 \right)^{1/4} \left( \hat b - \hat b^\dagger \right) + O\left(N^{-3/2}\right)                                                                                                                  \\
		\hat{S}_z' & = \frac{N}{2} + \frac{1}{2} \left(1 - \frac{1 + (B/J)^2}{\sqrt{1 + 2 (B/J)^2}}  \right) - \frac{1 + (B/J)^2}{\sqrt{1 + 2 (B/J)^2}} b^\dagger b + \frac{1}{2} \frac{(B/J)^2}{\sqrt{1 + 2 (B/J)^2}} \left(b^\dagger b^\dagger + b b \right)
	\end{aligned}
\end{equation}

Therefore, for the elements corresponding to diagonal terms in Bogoliubov basis we obtain,
\begin{equation}
	\begin{aligned}
		\frac{\langle n| \hat{S}_x' | n \rangle}{N}  & = \frac{\langle n| \hat{S}_y' |n\rangle}{N} = 0 \\
		\frac{\langle n | \hat{S}_z' | n \rangle}{N} & = \frac{1}{2} + \frac{f_s(n)}{N}
	\end{aligned}
\end{equation}
where
\begin{equation}
	f_s(n) = \frac{1}{2} \left(1 - \frac{1 + (B/J)^2}{\sqrt{1 + 2 (B/J)^2}}  \right) - n \frac{1 + (B/J)^2}{\sqrt{1 + 2 (B/J)^2}} = -n - \frac{1}{2}\left(\frac{1}{2} + n\right)\left(B/J\right)^4 +\left(\frac{1}{2} + n\right) \left(B/J\right)^6  + O\left(\left(B/J\right)^8\right)
\end{equation}

Moreover, for the first non-diagonal elements:
\begin{equation}
	\begin{aligned}
		\frac{\langle n - 1| \hat{S}_x' |n \rangle}{N} =\frac{\langle n| \hat{S}_x' |n - 1\rangle}{N}     & = \frac{1}{2} \left(1 + 2 (B/J)^2 \right)^{-1/4}\sqrt{\frac{n}{N}} + O\left(N^{-5/2}\right)  \\
		\frac{\langle n - 1| \hat{S}_y' |n \rangle}{N} = -\frac{\langle n| \hat{S}_y' |n - 1\rangle}{N}   & = \frac{1}{2i} \left(1 + 2 (B/J)^2 \right)^{1/4}\sqrt{\frac{n}{N}}  + O\left(N^{-5/2}\right) \\
		\frac{\langle n-1 | \hat{S}_z' | n \rangle}{N} = \frac{\langle n | \hat{S}_z' | n - 1 \rangle}{N} & = 0
	\end{aligned}
\end{equation}

It is important to note that the diagonal contribution vanishes due to the presence of an odd number of creation and annihilation operators in the expectation values of \(\hat{S}_x\) and \(\hat{S}_y\), in contrast to \(\hat{S}_z\). This asymmetry, arising from their respective operator orders, can be properly accounted for through a rotation of the spin components via the unitary transformation \(\hat{U}_R\), yielding:
\begin{equation}\label{eq:sm-average-magnetization}
	\begin{aligned}
		\hat{S}_x & = \sqrt{1 - \frac{B^2}{J^2}}\, \hat{S}_x' + \frac{B}{J}\, \hat{S}_z', \\
		\hat{S}_y & = \hat{S}_y',                                                         \\
		\hat{S}_z & = \sqrt{1 - \frac{B^2}{J^2}}\, \hat{S}_z' - \frac{B}{J}\, \hat{S}_x'.
	\end{aligned}
\end{equation}

The matrix elements corresponding to \(n\)-state approximately (considering \(B/J \ll 1\)) are
\begin{equation}
	\begin{aligned}
		\frac{1}{N} \bra{n} {\hat{S}_x} \ket{n} & = \frac{1}{2}  \frac{B}{J} \left(\frac{-\left(\left(B/J\right)^2 (2 n+1)\right)+\sqrt{2 \left(B/J\right)^2+1}-2
		n-1}{ N\sqrt{2 \left(B/J\right)^2+1}}+1\right)                                                                                                                                                                                              \\
		                                        & = \left( \frac{1}{2} - \frac{n}{N} \right) \frac{B}{J}  + \mathcal{O}\left( \frac{B^5}{J^5} \right)                                                                                               \\
		\frac{1}{N} \bra{n} {\hat{S}_y} \ket{n} & = 0                                                                                                                                                                                               \\
		\frac{1}{N} \bra{n} {\hat{S}_z} \ket{n} & =
		\frac{1}{2} \sqrt{1-\left(B/J\right)^2} \left(
		\frac{-\left(\left(B/J\right)^2 (2n+1)\right)+\sqrt{2 \left(B/J\right)^2+1}-2 n-1}{N\sqrt{2 \left(B/J\right)^2+1}}+1
		\right)
		\\
		                                        & =\left( \frac{1}{2} - \frac{n}{N} \right) \left(1 - \frac{1}{2}\frac{B^2}{J^2}  - \frac{1}{8}\frac{B^4}{J^4}\right) - \frac{2n+1}{4N}\frac{B^4}{J^4}  + \mathcal{O}\left( \frac{B^6}{J^6} \right)
	\end{aligned}
\end{equation}

The matrix element in a quasi-degenerate doublet—constructed as a cat-like superposition of symmetry-broken states \(\ket{\Uparrow_n}\) and \(\ket{\Downarrow_n}\)—can be approximated semiclassically. Since \(\ket{\Uparrow_n} \propto \ket{n}\) in the eigen spectrum, the off-diagonal matrix element between the conjugate eigenstates \(\ket{E_n} = \frac{1}{\sqrt{2}} \left( \ket{\Uparrow_n} + p_n \ket{\Downarrow_n} \right)\) and \(\ket{E_{\bar{n}}} = \frac{1}{\sqrt{2}} \left( \ket{\Uparrow_n} - p_n \ket{\Downarrow_n} \right)\) reads:
\begin{equation}
	\begin{aligned}
		O_{n\bar{n}} & = \bra{E_n} \hat{S}_z \ket{\bar{E}_{n}}                                                                                                                                                                                                  \\
		             & = \frac{1}{2} \left( \bra{\Uparrow_n} + p_n \bra{\Downarrow_n} \right) \hat{S}_z \left( \ket{\Uparrow_n} - p_n \ket{\Downarrow_n} \right)                                                                                                \\
		             & = \frac{1}{2} \left( \bra{\Uparrow_n} \hat{S}_z \ket{\Uparrow_n} - p_n \bra{\Uparrow_n} \hat{S}_z \ket{\Downarrow_n} + p_n \bra{\Downarrow_n} \hat{S}_z \ket{\Uparrow_n} - p_n^2 \bra{\Downarrow_n} \hat{S}_z \ket{\Downarrow_n} \right) \\
		             & =  \left( \bra{\Uparrow_n} \hat{S}_z \ket{\Uparrow_n} + \frac{p_n}{2} \bra{\Uparrow_n} \hat{X} \hat{S}_z - \hat{S}_z \hat{X} \ket{\Uparrow_n} \right)                                                                                    \\
		             & \simeq  \bra{n} {\hat{S}_z} \ket{n}
	\end{aligned}
\end{equation}
where we used the properties of Eq.~\eqref{eq:anticomm-x-sz}, namely that \(\hat X \ket{\Uparrow_n} = \ket{\Downarrow_n}\), \(\hat{X} \hat{S}_z\hat{X}=-\hat{S}_z\) and the structure of the cat states from Eq.~\eqref{eq:sm-cat-states-def}. In particular, in order to compute the term \( \langle \Uparrow_n | \hat{X} \hat{S}_z | \Uparrow_n \rangle \) in the last line, one can first apply  \(\hat{X} \hat{S}_z\) to \(|\Uparrow_n\rangle\) and then using \(\hat{X} |\lambda_j\rangle = |-\lambda_j\rangle\) one obtains that,
\begin{equation}
	\hat{X} \hat{S}_z |\Uparrow_n\rangle = \hat{X} \sum_{j,\,\lambda_j>0} c_{nj}^> \lambda_j |\lambda_j\rangle
	= \sum_{j,\,\lambda_j>0} c_{nj}^> \lambda_j |-\lambda_j\rangle.
\end{equation}
Taking the inner product with \(\langle \Uparrow_n |\) yields,
\begin{equation}
	\langle \Uparrow_n | \hat{X} \hat{S}_z |\Uparrow_n \rangle
	= \sum_{k,j,\,\lambda_{k,j}>0} (c_{nk}^>)^* c_{nj}^> \lambda_j \langle \lambda_k | -\lambda_j \rangle.
\end{equation}
By the same argument, one can also find that \(\langle \Uparrow_n | \hat{S}_z \hat{X} | \Uparrow_n \rangle = \sum_{k,j,\,\lambda_{k,j}>0} (c_{nk}^>)^* c_{nj}^> (-\lambda_j) \langle \lambda_k | -\lambda_j \rangle\) and we see that
\begin{equation}
	\bra{\Uparrow_n} \hat{X} \hat{S}_z - \hat{S}_z \hat{X} \ket{\Uparrow_n}  = 2 \sum_{k,j,\,\lambda_{k,j}>0} {\lambda_j (c_{nk}^>)^* c_{nj}^>  \langle \lambda_k | -\lambda_j \rangle} = 0
\end{equation}
where the result vanishes due to the orthogonality of eigenvectors, \(\langle \lambda_k | -\lambda_j \rangle = 0\), for eigenvalues with opposite signs.

\section{QFI dynamics in the LMG model}
\label{sec:sm-qfi-lmg-dynamics}
\begin{figure}
	\includegraphics[width=\textwidth]{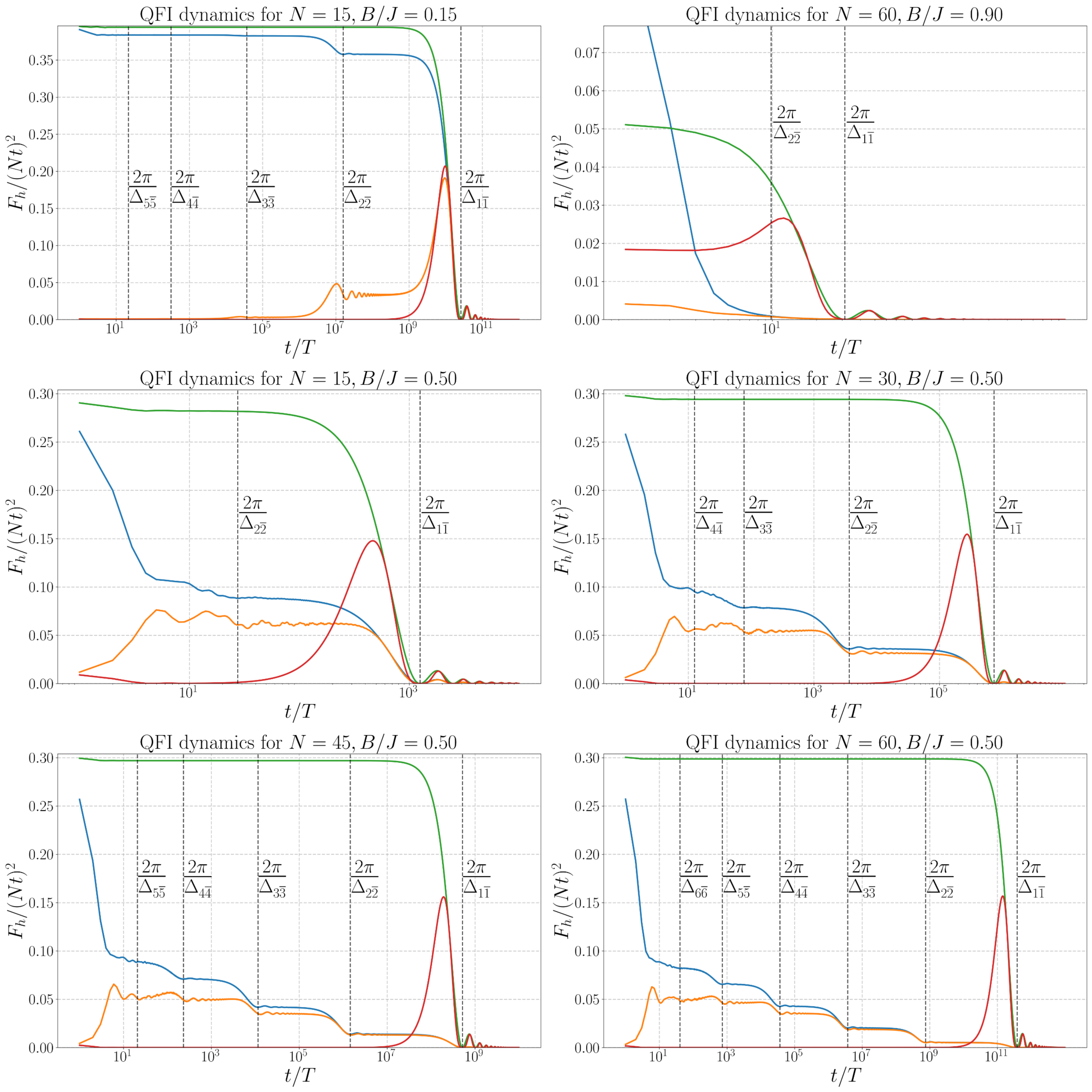}
	\caption{ Dynamics of QFI in the FTC sensor based on the LMG model, for different initial preparations, varying system size \(N\) and ratio \(B/J\). The simulations include four initial states: (green) a system prepared in the ground state of the Floquet Hamiltonian, \(|E_1\rangle\); (red) in a superposition of \(\pi\)-paired eigenstates, \((|E_1\rangle + |E_{\bar{1}}\rangle)/\sqrt{2}\); (orange) in an initial state with all spins aligned in the up direction, \(|\uparrow \dots \uparrow\rangle\); or (blue) a highly correlated initial state with a superposition of up and down spins, \( (|\uparrow \dots \uparrow\rangle + |\downarrow \dots \downarrow\rangle)/\sqrt{2}\).
	}
	\label{fig:sm-qfi-lmg-params}
\end{figure}

In this section we show an extended analysis of the QFI in the LMG-FTC, studying its dynamics for different system sizes \(N\) and transverse field \(B/J\) (spanning the FTC phase below the critical point \(B/J = 1\)).
These results, shown in Fig.~\ref{fig:sm-qfi-lmg-params}, corroborate the key qualitative features - Heisenberg scaling, \(F_h(t)\sim N^2 t^2\), sustained up to exponentially long times in \(N\), and step-like QFI plateaus arising from decoherence along cat subspaces - which persist across a range of parameters without fine-tuning.
We remark that in our simulations, the dynamics for large systems or those with small energy gaps (\(\Delta_{i\bar{i}} \approx 10^{-15}\)) prove challenging without high-precision numerical packages. We use MPmath for this task~\cite{TsypilnikovLMGCode2025} and observe an agreement between the analytical predictions and numerical results.

\end{document}